\documentclass{article}
\usepackage[preprint]{neurips_data_2021}

\usepackage[utf8]{inputenc} 
\usepackage[T1]{fontenc}    
\usepackage[hyphens]{url}
\usepackage{hyperref}
\hypersetup{breaklinks=true}
\usepackage{booktabs}       
\usepackage{amsfonts}       
\usepackage{nicefrac}       
\usepackage{microtype}      
\usepackage[dvipsnames,table,xcdraw]{xcolor}         
\usepackage[group-separator={,},group-four-digits]{siunitx}
\usepackage{listings}
\usepackage{pdfpages}

\usepackage{adjustbox}
\usepackage{graphicx}

\usepackage[font={small,it}]{caption}
\usepackage{subfig}
\usepackage{wrapfig}
\usepackage[nomarkers,tablesonly]{endfloat}

\usepackage{graphicx}
\usepackage{amsmath}

\usepackage{titlesec}
\titlespacing{\section}{0pt}{0.5ex}{0.5ex}
\titlespacing{\subsection}{0pt}{0ex}{0ex}
\titlespacing{\subsubsection}{0pt}{0.5ex}{0ex}

\usepackage{booktabs,dcolumn}
\newcolumntype{d}[1]{D..{#1}}
\newcommand{\rCE}{\textsf{$rCE$}}
\newcommand{\mrCE}{\textsf{$mrCE$}}

\title{Robustness Disparities in Commercial Face Detection}

\author{%
  Samuel Dooley \\
  University of Maryland\\
  \texttt{sdooley1@cs.umd.edu} \\
   \And
   Tom Goldstein \\
  University of Maryland\\
  \texttt{tomg@cs.umd.edu} \\
   \And
   John P. Dickerson \\
  University of Maryland\\
  \texttt{john@cs.umd.edu} \\
}

\newcommand{\FRT}{FRT}

\begin{document}

\maketitle

\begin{abstract}
  Facial detection and analysis systems have been deployed by large companies and critiqued by scholars and activists for the past decade. Critiques that focus on system performance analyze disparity of the system's output, i.e., how frequently is a face detected for different Fitzpatrick skin types or perceived genders. However, we focus on the robustness of these system outputs under noisy natural perturbations. We present the first of its kind detailed benchmark of the robustness of three such systems: Amazon Rekognition, Microsoft Azure, and Google Cloud Platform.  We use both standard and recently released academic facial datasets to quantitatively analyze trends in robustness for each. Across all the datasets and systems, we generally find that photos of individuals who are \emph{older}, \emph{masculine presenting}, of \emph{darker skin type}, or have \emph{dim lighting} are more susceptible to errors than their counterparts in other identities.
\end{abstract}

\section{Introduction}\label{sec:intro}
Face detection systems identify the presence and location of faces in images and video.  Automated face detection is a core component of myriad systems---including \emph{face recognition technologies} (\FRT{}), wherein a detected face is matched against a database of faces, typically for identification or verification purposes.  \FRT{}-based systems are widely deployed~\citep{hartzog2020secretive, derringer2019surveillance, weise2020amazon}.  Automated face recognition enables capabilities ranging from the relatively morally neutral (e.g., searching for photos on a personal phone~\citep{GooglePhotosFRT}) to morally laden (e.g., widespread citizen surveillance~\citep{hartzog2020secretive}, or target identification in warzones~\citep{marson_forrest_2021}).  Legal and social norms regarding the usage of \FRT{} are evolving~\citep[e.g.,][]{grother2019face}. For example, in June 2021, the first county-wide ban on its use for policing~\cite[see, e.g.,][]{garvie2016perpetual} went into effect in the US~\citep{Gutman21:King}.  Some use cases for \FRT{} will be deemed socially repugnant and thus be either legally or \emph{de facto} banned from use; yet, it is likely that pervasive use of facial analysis will remain---albeit with more guardrails than are found today~\citep{singer2018microsoft}.

One such guardrail that has spurred positive, though insufficient, improvements and widespread attention is the use of benchmarks. For example, in late 2019, the US National Institute of Standards and Technology (NIST) adapted its venerable Face Recognition Vendor Test (FRVT) to explicitly include concerns for demographic effects~\citep{grother2019face}, ensuring such concerns propagate into industry systems.   Yet, differential treatment by \FRT{} of groups has been known for at least a decade~\citep[e.g.,][]{klare2012face,el2016face}, and more recent work spearheaded by~\citet{buolamwini2018gendershades} uncovers unequal performance at the phenotypic subgroup level.  That latter work brought widespread public, and thus burgeoning regulatory, attention to bias in \FRT{}~\citep[e.g.,][]{lohr2018facial,CodedBias}.

One yet unexplored benchmark examines the bias present in a system's robustness (e.g., to noise, or to different lighting conditions), both in aggregate and with respect to different dimensions of the population on which it will be used.
Many detection and recognition systems are not built in house, instead making use of commercial cloud-based ``ML as a Service'' (MLaaS) platforms offered by tech giants such as Amazon, Microsoft, Google, Megvii, etc.  The implementation details of those systems are not exposed to the end user---and even if they were, quantifying their failure modes would be difficult.  With this in mind, our \textbf{main contribution} is a wide \emph{robustness benchmark} of three commercial-grade face detection systems (accessed via Amazon's Rekognition, Microsoft's Azure, and Google Cloud Platform's face detection APIs).  For fifteen types of realistic noise, and five levels of severity per type of noise~\citep{hendrycks2019benchmarking}, we test both APIs against images in each of four well-known datasets.  Across these more than \num{5000000} noisy images, we analyze the impact of noise on face detection performance.  Perhaps unsurprisingly, we find that noise decreases overall performance, though the result from our study confirms the previous findings of~\citet{hendrycks2019benchmarking}. Further, different types of noise impact, in an unbiased way, cross sections of the population of images (e.g., based on Fitzgerald skin type, age, self-identified gender, and intersections of those dimensions).  Our method is extensible and can be used to quantify the robustness of other detection and \FRT{} systems, and adds to the burgeoning literature supporting the necessity of explicitly considering bias in ML systems with morally-laden downstream uses.

\section{Related Work}\label{sec:rel-work}

We briefly overview additional related work in the two core areas addressed by our benchmark: robustness to noise and demographic disparity in facial detection and recognition.  That latter point overlaps heavily with the fairness in machine learning literature; for additional coverage of that broader ecosystem and discussion around bias in machine learning writ large, we direct the reader to survey works due to~\citet{chouldechova2018frontiers} and~\citet{fairmlbook}.

\paragraph{Demographic effects in facial detection and recognition.}
The existence of differential performance of facial detection and recognition on groups and subgroups of populations has been explored in a variety of settings.  Earlier work~\citep[e.g.,][]{klare2012face,o2012demographic} focuses on single-demographic effects (specifically, race and gender) in pre-deep-learning face detection and recognition.  \citet{buolamwini2018gendershades} uncovers unequal performance at the phenotypic subgroup level in, specifically, a gender classification task powered by commercial systems.  That work, typically referred to as ``Gender Shades,'' has been and continues to be hugely impactful both within academia and at the industry level.  Indeed, \citet{raji2019actionable} provide a follow-on analysis, exploring the impact of the~\citet{buolamwini2018gendershades} paper publicly disclosing performance results, for specific systems, with respect to demographic effects; they find that their named companies (IBM, Microsoft, and Megvii) updated their APIs within a year to address some concerns that were surfaced.  Subsequently, the late 2019 update to the NIST FRVT provides evidence that commercial platforms are continuing to focus on performance at the group and subgroup level~\citep{grother2019face}. Further recent work explores these demographic questions with a focus on Indian election candidates \citep{jain2021Million}. We see our benchmark as adding to this literature by, for the first time, addressing both noise and demographic effects on commercial platforms' face detection offerings.

In this work, we focus on \emph{measuring} the impact of noise on a classification task,  like that of \citet{wilber2016can}; indeed, a core focus of our benchmark is to \emph{quantify} relative drops in performance conditioned on an input datapoint's membership in a particular group.  We view our work as a \emph{benchmark}, that is, it focuses on quantifying and measuring, decidedly not providing a new method to ``fix'' or otherwise mitigate issues of demographic inequity in a system.  Toward that latter point, existing work on ``fixing'' unfair systems can be split into three (or, arguably, four~\citep{savani2020posthoc}) focus areas: pre-, in-, and post-processing. Pre-processing work largely focuses on dataset curation and preprocessing~\citep[e.g.,][]{Feldman2015Certifying, ryu2018inclusivefacenet, quadrianto2019discovering, wang2020mitigating}. In-processing often constrains the ML training method or optimization algorithm itself~\citep[e.g.,][]{zafar2017aistats, Zafar2017www, zafar2019jmlr, donini2018empirical, goel2018non,Padala2020achieving, agarwal2018reductions, wang2020mitigating,martinez2020minimax,diana2020convergent,lahoti2020fairness}, or focuses explicitly on so-called fair representation learning~\citep[e.g.,][]{adeli2021representation,dwork2012fairness,zemel13learning,edwards2016censoring,madras2018learning,beutel2017data,wang2019balanced}. Post-processing techniques adjust decisioning at inference time to align with quantitative fairness definitions~\citep[e.g.,][]{hardt2016equality,wang2020fairness}.

\paragraph{Robustness to noise.}

Quantifying, and improving, the robustness to noise of face detection and recognition systems is a decades-old research challenge.  Indeed, mature challenges like NIST's Facial Recognition Vendor Test (FRVT) have tested for robustness since the early 2000s~\citep{phillips2007frvt}.  We direct the reader to a comprehensive introduction to an earlier robustness challenge due to NIST~\citep{phillips2011introduction}; that work describes many of the specific challenges faced by face detection and recognition systems, often grouped into Pose, Illumination, and Expression (PIE).  It is known that commercial systems still suffer from degradation due to noise~\citep[e.g.,][]{hosseini2017google}; none of this work also addresses the intersection of noise with bias, as we do.
Recently, \emph{adversarial} attacks have been proposed that successfully break commercial face recognition systems~\citep{shan2020fawkes,cherepanova2021lowkey}; we note that our focus is on \emph{natural} noise, as motivated by~\citet{hendrycks2019benchmarking} by their ImageNet-C benchmark.  Literature at the intersection of adversarial robustness and fairness is nascent and does not address commercial platforms~\citep[e.g.,][]{singh2020robustness,nanda2021fairness}.  To our knowledge, our work is the first systematic benchmark for commercial face detection systems that addresses, comprehensively, noise and its differential impact on (sub)groups of the population.

\section{Experimental Description}\label{sec:experimental-design}

\paragraph{Datasets and Protocol.}
\begin{figure}
    \centering
    \includegraphics[width=\textwidth]{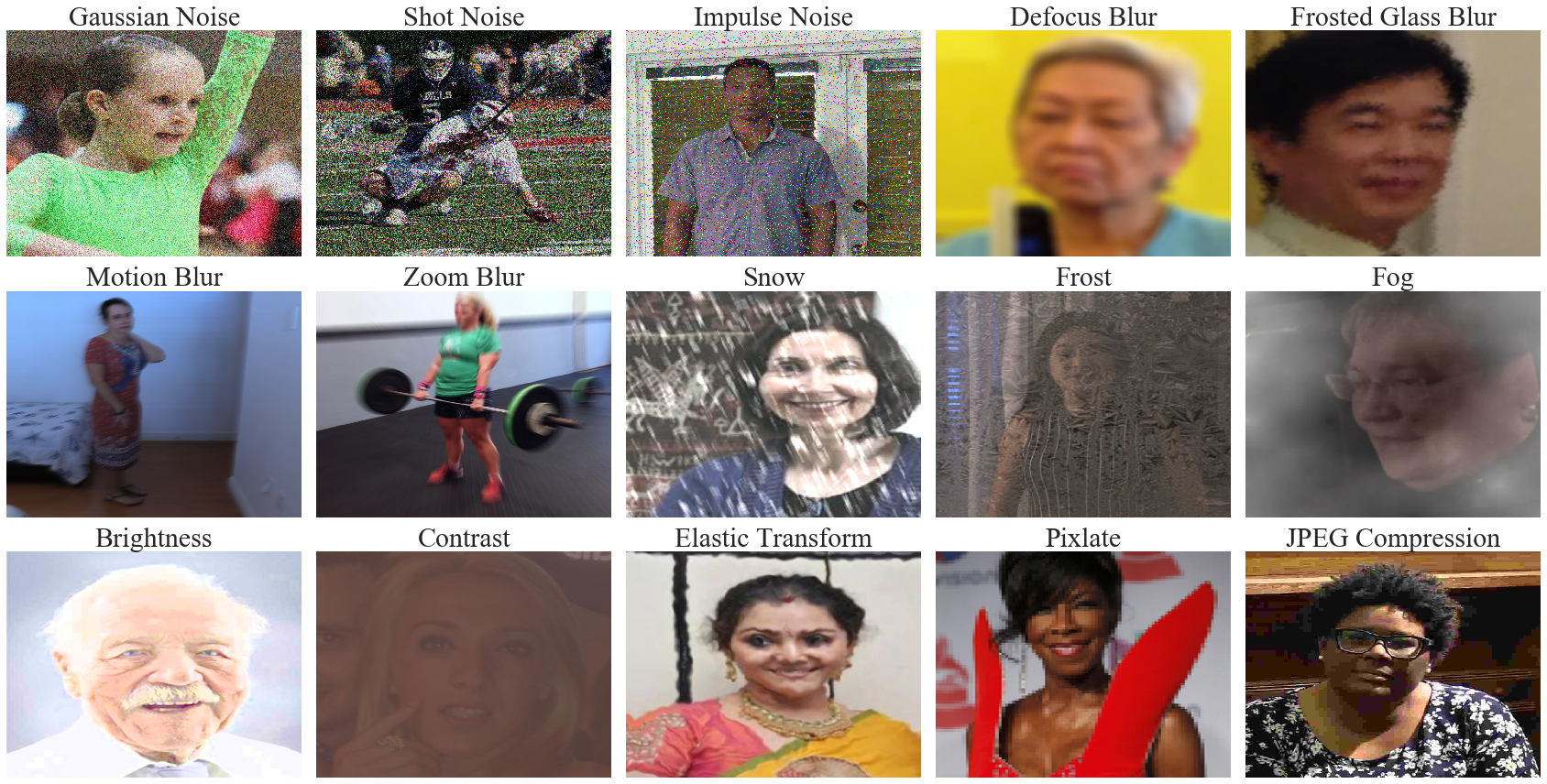}
    \caption{Our benchmark consists of 5,066,312 images of the 15 types of algorithmically generated corruptions produced by ImageNet-C. We use data from four datasets (Adience, CCD, MIAP, and UTKFace) and present examples of corruptions from each dataset here.}
    \label{fig:examples}
\end{figure}

This benchmark uses four datasets to evaluate the robustness of Amazon AWS and Microsoft Azure's face detection systems. They are described below and a repository for the experiments can be found here: \url{https://github.com/dooleys/Robustness-Disparities-in-Commercial-Face-Detection}.

The Open Images Dataset V6 -- Extended; More Inclusive Annotations for People ({\bf MIAP}) dataset \citep{miap} was released by Google in May 2021 as a  extension of the popular, permissive-licensed Open Images Dataset specifically designed to improve annotations of humans. For each image, every human is exhaustively annotated with bounding boxes for the entirety of their person visible in the image. Each annotation also has perceived gender (Feminine/Masculine/Unknown) presentation and perceived age (Young, Middle, Old, Unknown) presentation.   

The Casual Conversations Dataset ({\bf CCD}) \citep{ccd} was released by Facebook in April 2021 under limited license and includes videos of actors. Each actor consented to participate in an ML dataset and provided their self-identification of age and gender (coded as Female, Male, and Other), each actor's skin type was rated on the Fitzpatrick scale \citep{fitzpatrick1988validity}, and each video was rated for its ambient light quality. For our benchmark, we extracted one frame from each video. 

The {\bf Adience} dataset \citep{adience} under a CC license, includes cropped images of faces from images ``in the wild''. Each cropped image contains only one primary, centered face, and each face is annotated by an external evaluator for age and gender (Female/Male). The ages are reported as member of 8 age range buckets: 0-2; 3-7; 8-14; 15-24; 25-35; 36-45; 46-59; 60+.  

Finally, the {\bf UTKFace} dataset \citep{utk} under a non-commercial license, contains images with one primary subject and were annotated for age (continuous), gender (Female/Male), and ethnicity (White/Black/Asian/Indian/Others) by an algorithm, then checked by human annotators.

For each of the datasets, we randomly selected a subset of images for our evaluation. We capped the number of images from each intersectional identity at \num{1500} as an attempt to reduce the effect of highly imbalanced datasets. We include a total of \num{66662} images with \num{14919} images from Adience; \num{21444} images from CCD; \num{8194} images from MIAP; and \num{22105} images form UTKFace. The full breakdown of totals of images from each group can be found in Section~\ref{sec:image_counts}.

Each image was corrupted a total of 75 times, per the ImageNet-C protocol with the main 15 corruptions each with 5 severity levels. Examples of these corruptions can be seen in Figure~\ref{fig:examples}. This resulted in a total of \num{5066312} images  (including the original clean ones) which were each passed through the AWS, Azure, and Google Cloud Platform (GCP) face analysis systems. A detailed description of which API settings were selected can be found in Appendix~\ref{sec:api}. The API calls were conducted between 19 May and 29 May 2021. Images were processed and stored within AWS's cloud using S3 and EC2. The total cost of the experiments was \$\num{17507.55} and a breakdown of costs can be found in Appendix~\ref{sec:costs}.

\paragraph{Evaluation Metrics.}

We evaluate the error of the face systems. Since none of the chosen datasets have ground truth face bounding boxes, we compare the number of detected faces from the clean image to the number of faces detected in a corrupted image, using the former as ground truth.

Our main metric is the relative error in the number of faces a system detects after corruption; this metric has been used in other facial processing benchmarks \citep{jain2021Million}. Measuring error in this way is in some sense incongruous with the object detection nature of the APIs. However, none of the data in our datasets have bounding boxes for each face. This means that we cannot calculate precision metrics as one would usually do with other detection tasks. To overcome this, we hand-annotated bounding boxes for each face in 772 (1.2\% of the dataset) random images from the dataset. We then calculated per-image precision scores (with an intersection over union of 0.5) and per-image relative error in face counts and we find a Pearson's correlation of 0.91 (with $p<0.001$). This high correlation indicates that the proxy is sufficient to be used in this benchmark in the absence of fully annotated bounding boxes.

This error is calculated for each image. Specifically, we first pass every clean, uncorrupted image through the commercial system's API. Then, we measure the number of detected faces, i.e., length of the system's response, and treat this number as the ground truth. Subsequently, we compare the  number of detected faces for a corrupted version of that image. If the two face counts are not the same, then we call that an error. We refer to this as the \emph{relative corruption error}. For each clean image, $i$, from dataset $d$, and each corruption $c$ which produces a corrupted image $\hat{i}_{c,s}$ with severity $s$, we compute the relative corruption error for system $r$ as

\[
    \rCE_{c,s}^{d,r}(\hat{i}_{c,s}) :=
        \begin{cases}
        1, \quad \text{if }  l_r(i) \neq l_r(\hat{i}_{c,s})\\ 
        0, \quad \text{if }   l_r(i) = l_r(\hat{i}_{c,s})\\ 
        \end{cases}
\]
 where $l_r$ computes the number of detected faces, i.e., length of the response, from face detection system $r$ when given an image. Often the super- and subscripts are omitted when they are obvious. 

Our main metric, relative error, aligns with that of the ImageNet-C benchmark. We report mean relative corruption error ($\mrCE{}$) defined as taking the average of $\rCE{}$ across some relative set of categories. In our experiments, depending on the context, we might have any of the following categories: face systems, datasets, corruptions, severities, age presentation, gender presentation, Fitzpatrick rating, and ambient lighting. For example, we might report the relative mean corruption error when averaging across demographic groups; the mean corruption error for Azure on the UTK dataset for each age group $a$ is $\mrCE{}_a = \frac{1}{15}\frac{1}{5}\sum_{c,s} \rCE{}^{UTK,Azure}_{c,s,a}$. The subscripts on $\mrCE{}$ are omitted when it is obvious what their value is in whatever context they are presented.

Finally, we  also investigate the significance of whether the $\mrCE{}$ for two groups are equal. For example, our first question is whether the two commercial systems (AWS and Azure) have comparable $\mrCE{}$ overall. To do this, we report the raw $\mrCE{}$; these frequency or empiric probability statistics offer much insight into the likelihood of error. But we also indicate the statistical significance at $\alpha=0.05$ determined by logistic regressions for the appropriate variables and interactions. For each claim of significance, regression tables can be found in the appendix. Accordingly, we discuss the odds or odds ratio of relevant features. See Appendix~\ref{sec:metric_example} for a detailed example.  Finally, each claim we make for an individual dataset or service is backed up with statistical rigor through the logistic regressions. Each claim we make across datasets is done by looking at the trends in each dataset and are inherently qualitative.

\paragraph{What is not included in this study.}

There are three main things that this benchmark does not address. 
First, we do not examine cause and effect. We report inferential statistics without discussion of what generates them.
Second, we only examine the types of algorithmicaly generated natural noise present in the 15 corruptions. We speak narrowly about robustness to these corruptions or perturbations. We explicitly do not study or measure robustness to other types of changes to images, for instance adversarial noise, camera dimensions, etc. 
Finally, we do not investigate algorithmic training. We do not assume any knowledge of how the commercial system was developed or what training procedure or data were used. 

\paragraph{Social Context.}

The central analysis of this benchmark relies on socially constructed concepts of gender presentation and the related concepts of race and age. While this benchmark analyzes phenotypal versions of these from metadata on ML datasets, it would be wrong to interpret our findings absent a social lens of what these demographic groups mean inside a society. We guide the reader to~\citet{benthall2019racial} and \citet{hanna2020towards} for a look at these concepts for race in machine learning, and~\citet{hamidi2018gender} and \citet{keyes2018misgendering} for similar looks at gender. 

\section{Benchmark Results}
\begin{figure}
    \centering
    \includegraphics[width=\textwidth]{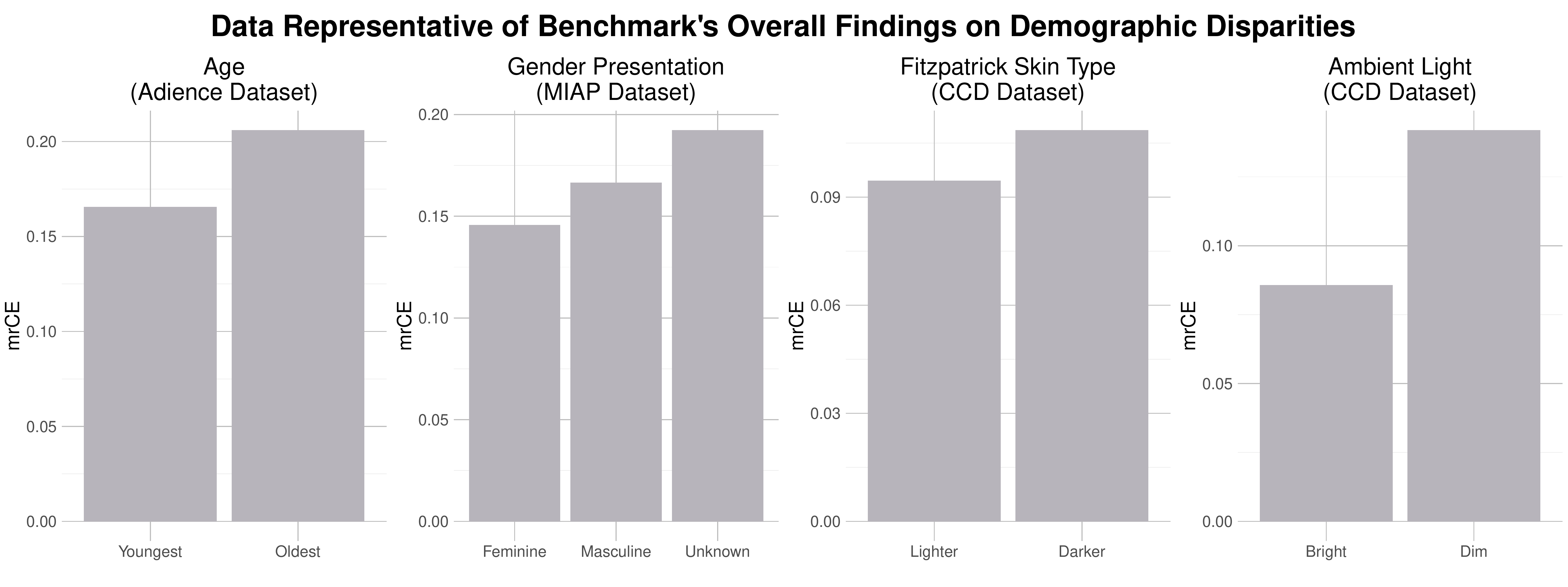}
    \caption{There are disparities in all of the demographics included in this study; we show representative evidence for each demographic on different datasets. On the left, we see (using Adience as an exemplar) that the  oldest two age groups are roughly 25\% more error prone than the  youngest two groups. Using MIAP as an exemplar, masculine presenting subjects are 20\% more error prone than feminine. On the CCD dataset, we find that individuals with Fitzpatrick scales IV-VI have a roughly 25\% higher chance of error than lighter skinned individuals. Finally, dimly lit individuals are 60\% more likely to have errors. }
    \label{fig:overall}
\end{figure}

We now report the main results of our benchmark, a synopsis of which is in Figure~\ref{fig:overall}. Our main results are derived from Table~\ref{tbl:comp_full} which report one regression for each dataset. Each regression includes all demographic variables and each variable is normalized for consistency across the datasets. Overall, we find that photos of individuals who are \emph{older}, \emph{masculine presenting}, \emph{darker skinned}, or are \emph{dimly lit} are more susceptible to errors than their counterparts. We see that each of these conclusions are consistent across datasets except that UTKFace has masculine presenting individuals as performing better than feminine presenting. 

\subsection{ System Performance}\label{sec:comp_system}

\begin{figure}
\centering
\begin{minipage}{.36\textwidth}
  \centering
  \includegraphics[width=\linewidth]{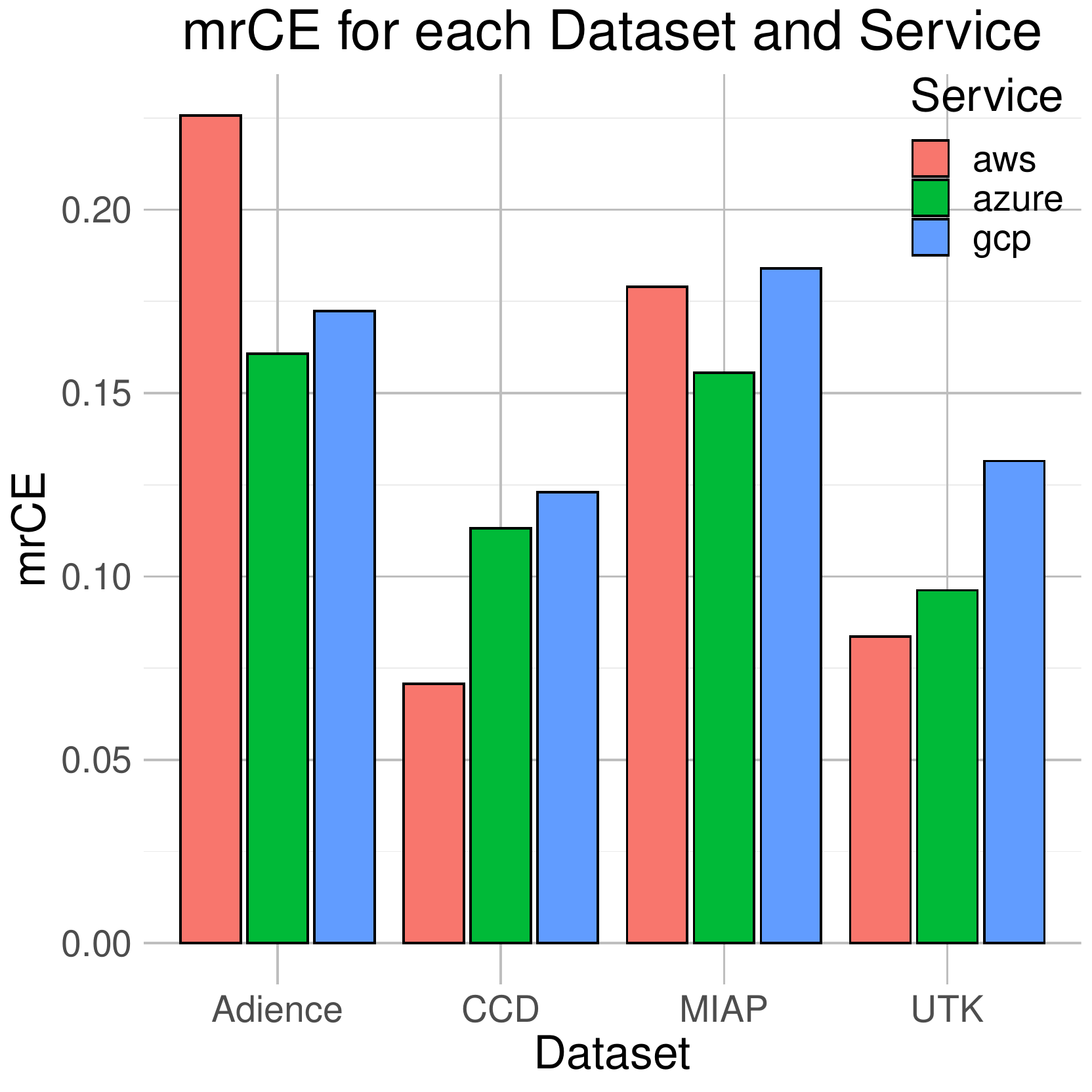}
  \captionof{figure}{Overall performance differences for each dataset and service}
  \label{fig:serice_comp}
\end{minipage}\hfill
\begin{minipage}{.6\textwidth}
  \centering
  \includegraphics[width=\linewidth]{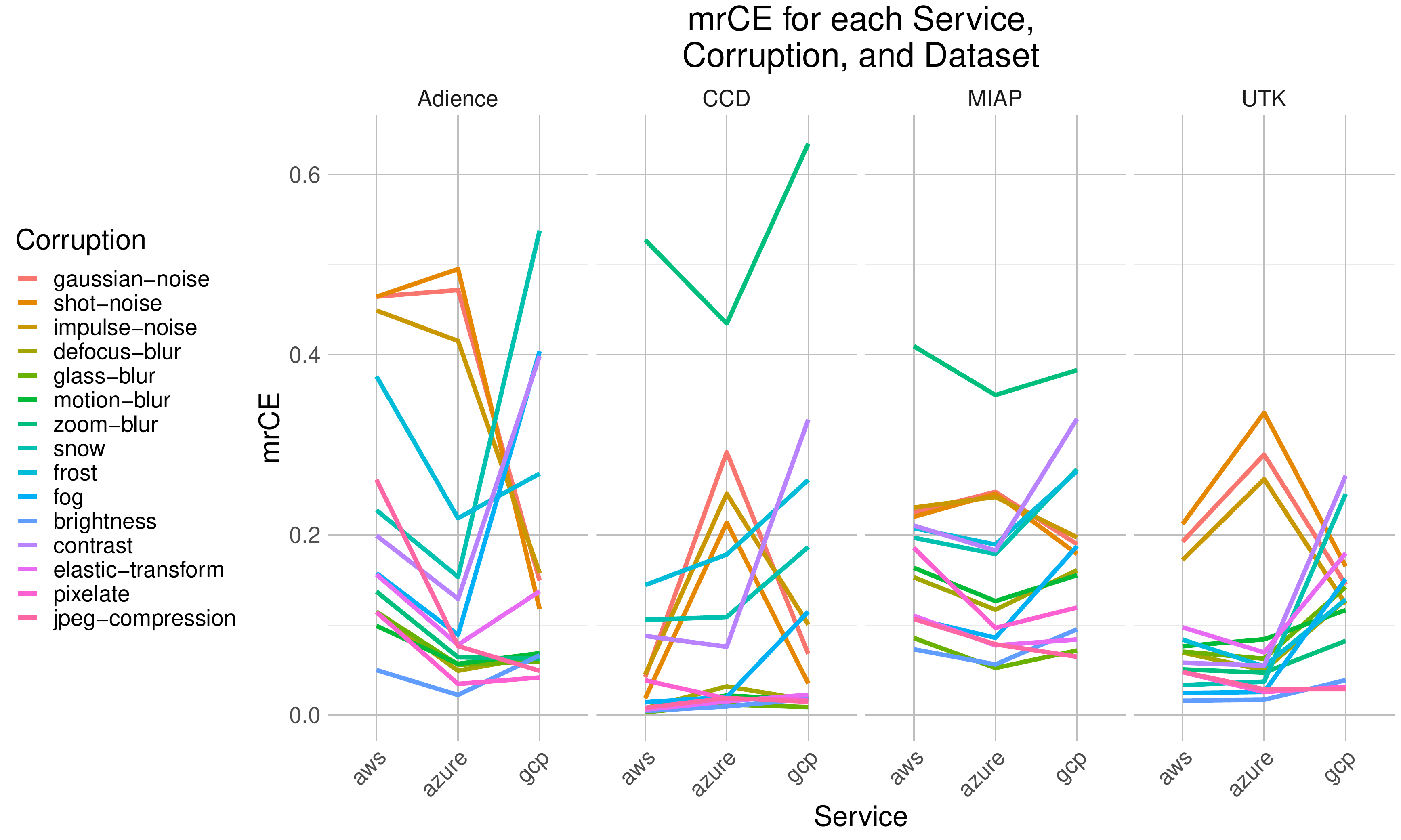}
  \captionof{figure}{A comparison of \mrCE{} for each corruption across each dataset and service.}
  \label{fig:corruption}
\end{minipage}
\end{figure}

We plot \mrCE{} for each dataset and service in Figure~\ref{fig:serice_comp}; the difference between services is statistically significant for each dataset and each service. The only consistent pattern is that GCP is always worse than Azure. 
These corruption errors should be compared to the clean errors which are very much less than 1\%. For comparison, the lowest \mrCE{} from Figure~\ref{fig:serice_comp} is 7\% for corrupted images. Thus, these corruptions are very significantly impacting the performance of these systems.

\subsection{Noise corruptions are the most difficult}\label{sec:comp_corr}

Recall that there are four types of ImageNet-C corruptions: noise, blur, weather, and digital. From Figure~\ref{fig:corruption}, we observe that the noise corruptions are markedly some of the most difficult corruptions for Azure to handle across the datasets, whereas GCP has better performance on noise corruptions than Azure and AWS. Though we can only stipulate, these differences might stem from pre-processing steps that each service takes before processing their image. GCP might have a robust noise pre-processing step, which would account for their superior performance with these corruptions. 

The zoom blur corruption proves particularly difficult on the CCD and MIAP datasets, though Azure is significantly better than AWS and GCP on both datasets. We also note that all corruptions for all datasets and commercial systems are significantly differently from zero. Further details can be found by examining Figure~\ref{fig:corruption} and Appendix Tables~\ref{tbl:comp_full}-\ref{tbl:comp_full_utk}.

\subsubsection{Comparison to ImageNet-C results}

We compare the \citet{hendrycks2019benchmarking} findings to our experiments. We recreate Figure 3 from their paper with more current results for recent models since their paper was published, as well as the addition of our findings; see Figure~\ref{fig:imagenetc_comp}. This figure reproduces their metric, mean corruption error and relative mean corruption error. From this figure, we can conclude that our results are very highly in-line with the predictions from the previous data. This indicates that, even with highly accurate models, accuracy is a strong predictor of robustness. 

We also examined the corruption-specific differences between our findings (with face data) and that of the original paper (with ImageNet data). We find that while ImageNet datasets are most susceptible to blurs and digital corruptions, facial datasets are most susceptible to noise corruptions, zoom blur, and weather. These qualitative differences deserve future study.

\subsection{Errors increase on older subjects}

\begin{figure}
\centering
  \includegraphics[width=.23\textwidth]{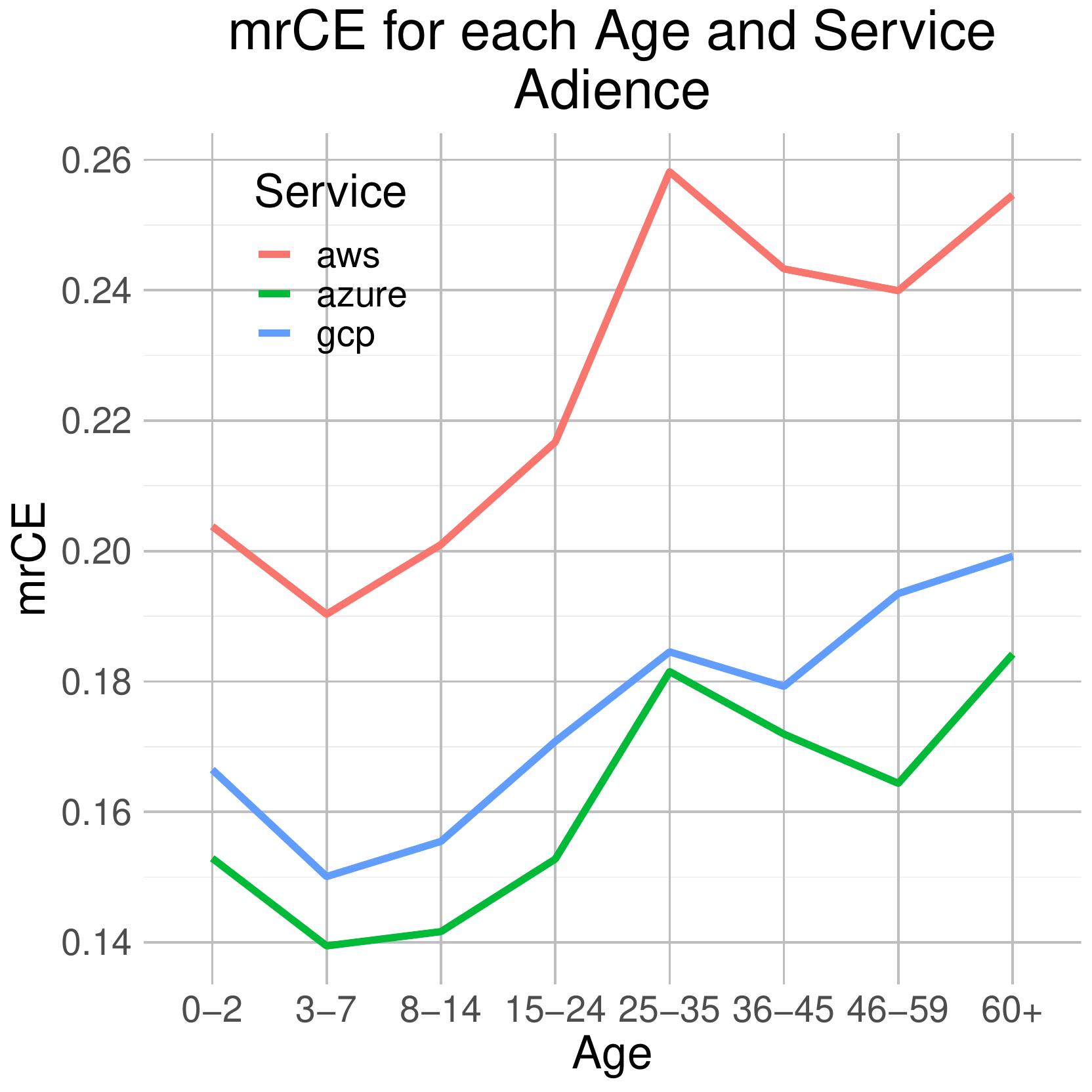}
  \includegraphics[width=.23\textwidth]{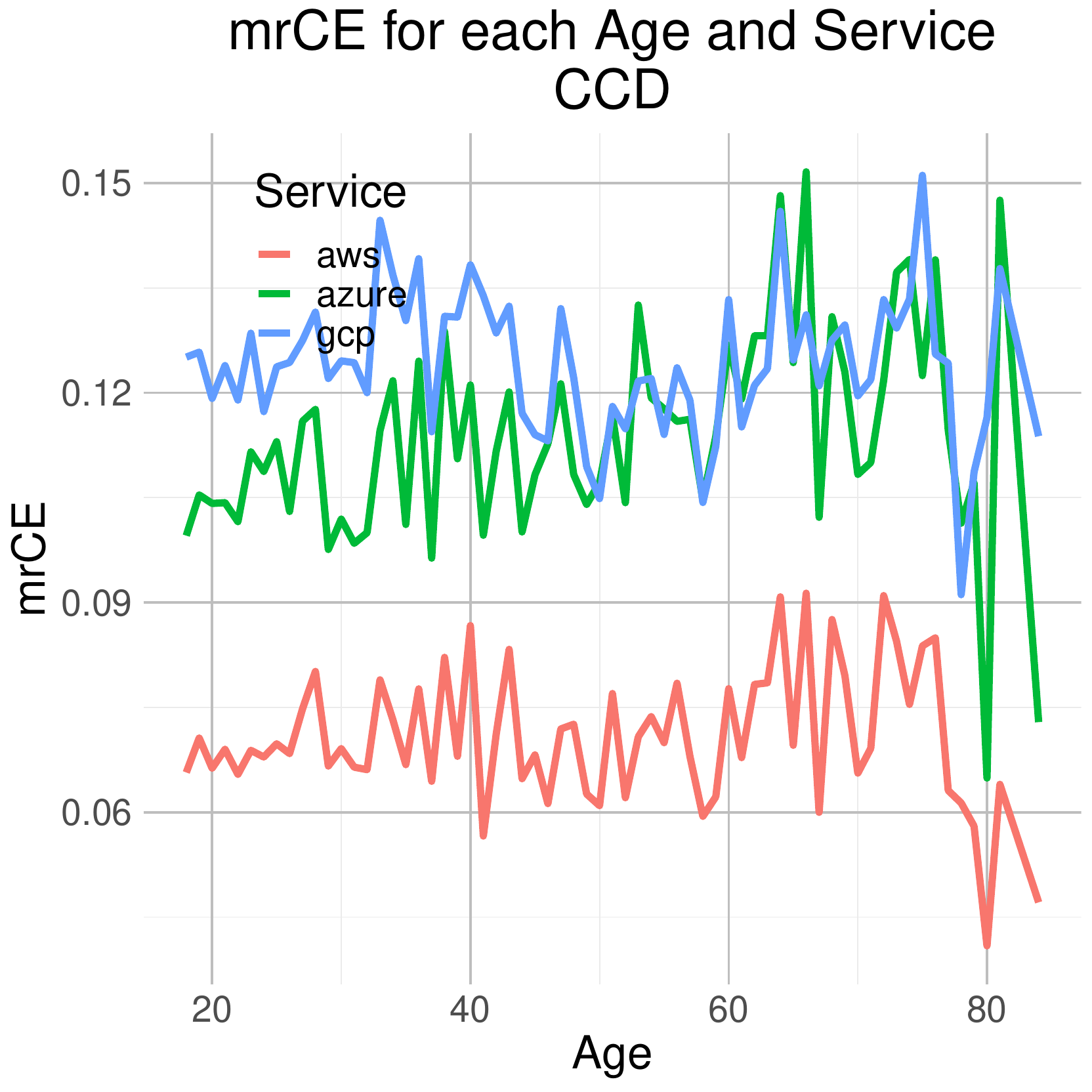}
  \includegraphics[width=.23\textwidth]{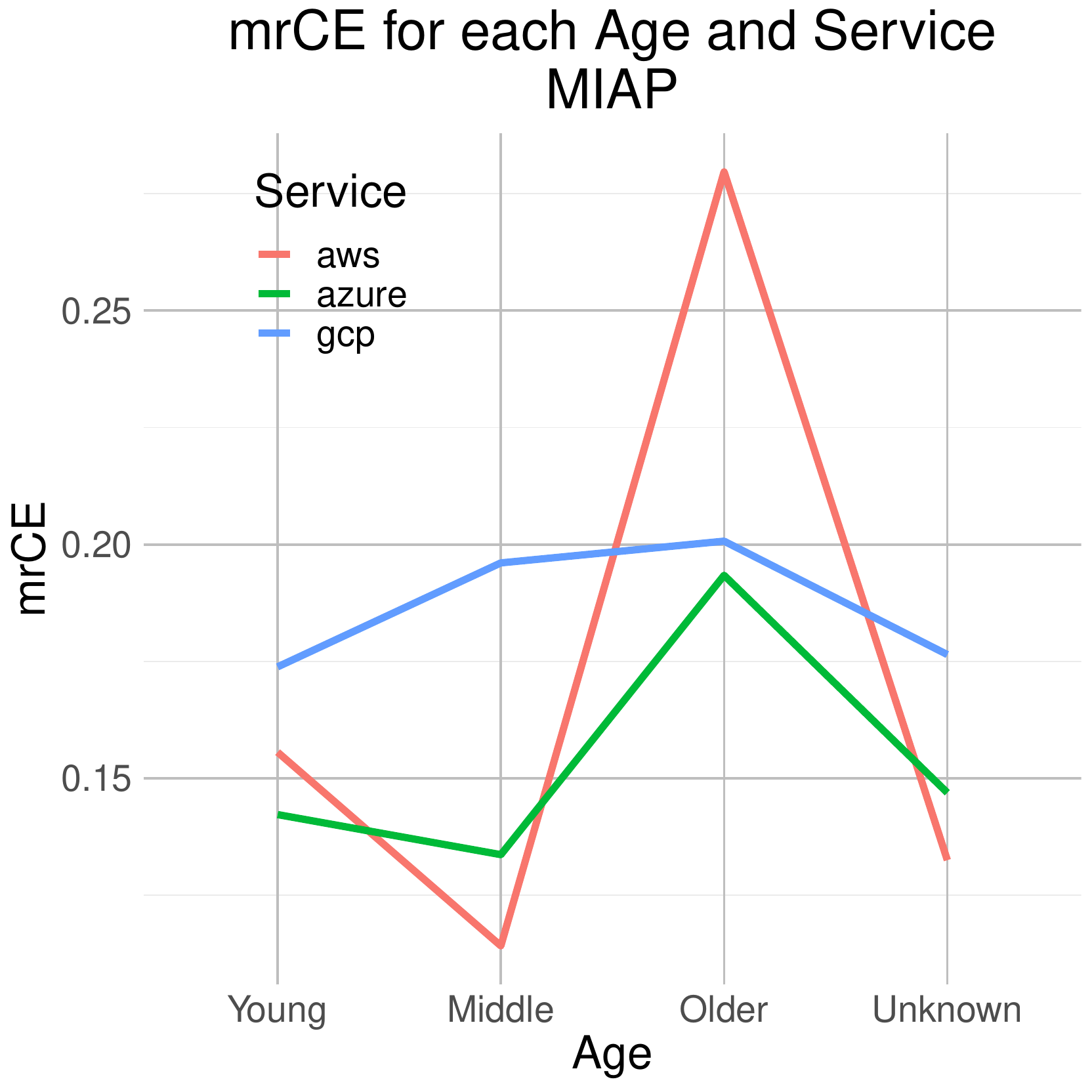}
  \includegraphics[width=.23\textwidth]{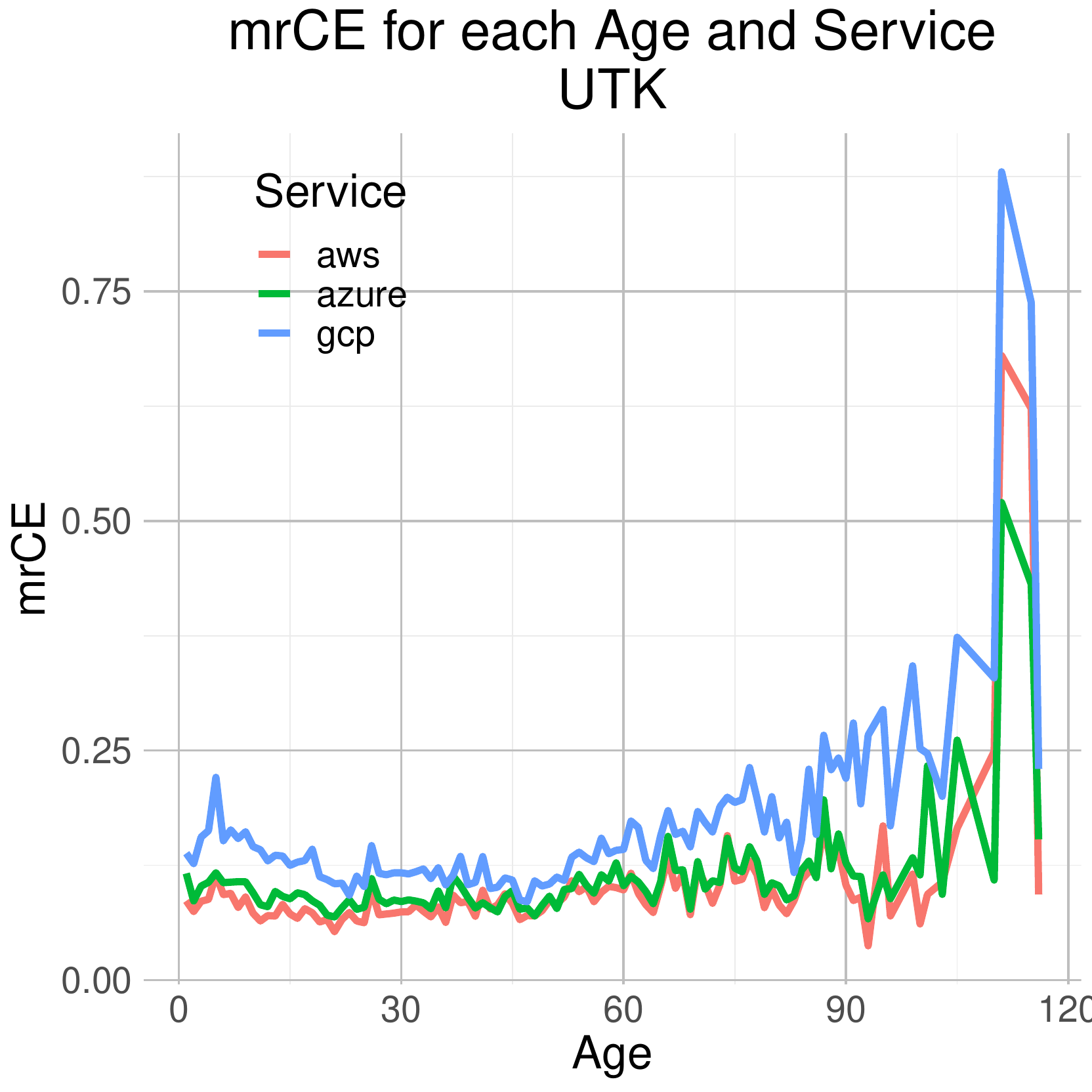}
\caption{Each figure depicts the \mrCE{} across ages. Each line depicts a commercial system. Age is a categorical variable for Adience and MIAP but a numeric for CCD and UTKFace.}
\label{fig:comp_age}
\end{figure}

We observe a significant impact of age on $\mrCE{}$; see Figure~\ref{fig:comp_age}. In every dataset and every commercial system (Appendix Tables~\ref{tbl:comp_full}-\ref{tbl:comp_full_utk}), we see that older subjects have significantly higher error rates. 

On the Adience dataset (Table~\ref{tbl:comp_full_adience}), the odds of error for the oldest group is 31\% higher than that of the youngest group. Interestingly, the shape of the \mrCE{} curves across the age groups is similar for each service. For the MIAP dataset (Table~\ref{tbl:comp_full_miap}), the age disparity is very pronounced. In AWS for instance, we see a 145\% increase in error for the oldest individuals. The overall odds ratio between the oldest and youngest is 1.383. 

The CCD and UTKFace datasets have numeric age. Analyzing the  regressions indicates that for every increase of 10 years, there is a 2.3\% increase in the likelihood of error on the CCD data and 2.7\% increase for UTKFace data.

\subsection{Masculine presenting individuals have more errors than feminine presenting}

\begin{figure}
\centering
\parbox{.29\textwidth}{%
\centering
  \includegraphics[width=.29\textwidth]{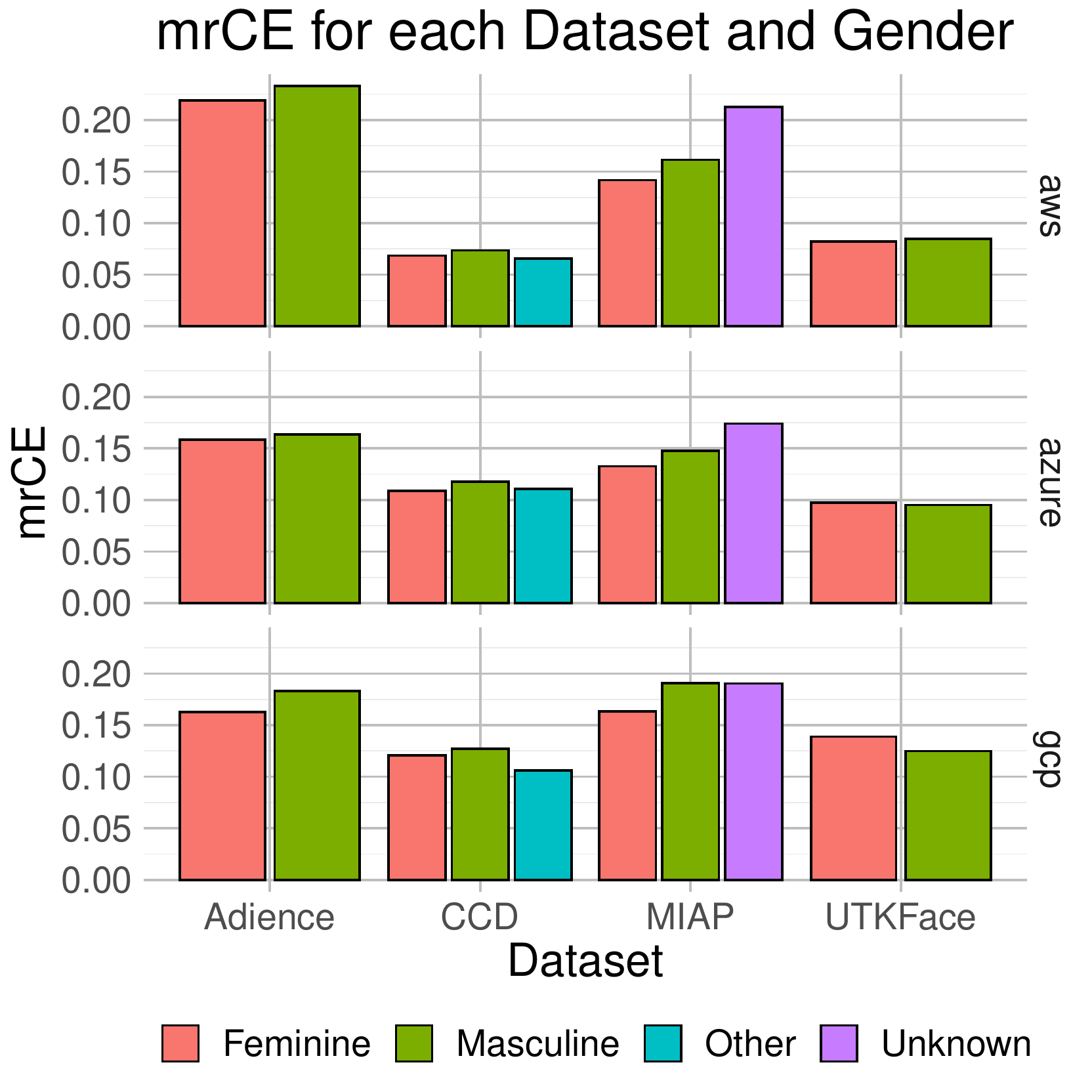}
\caption{Observe that on all datasets, except for UTKFace, feminine presenting individuals are more robust than masculine.}%
\label{fig:comp_gender}}%
\qquad
\begin{minipage}{.3\textwidth}%
  \centering
  \includegraphics[width=\textwidth]{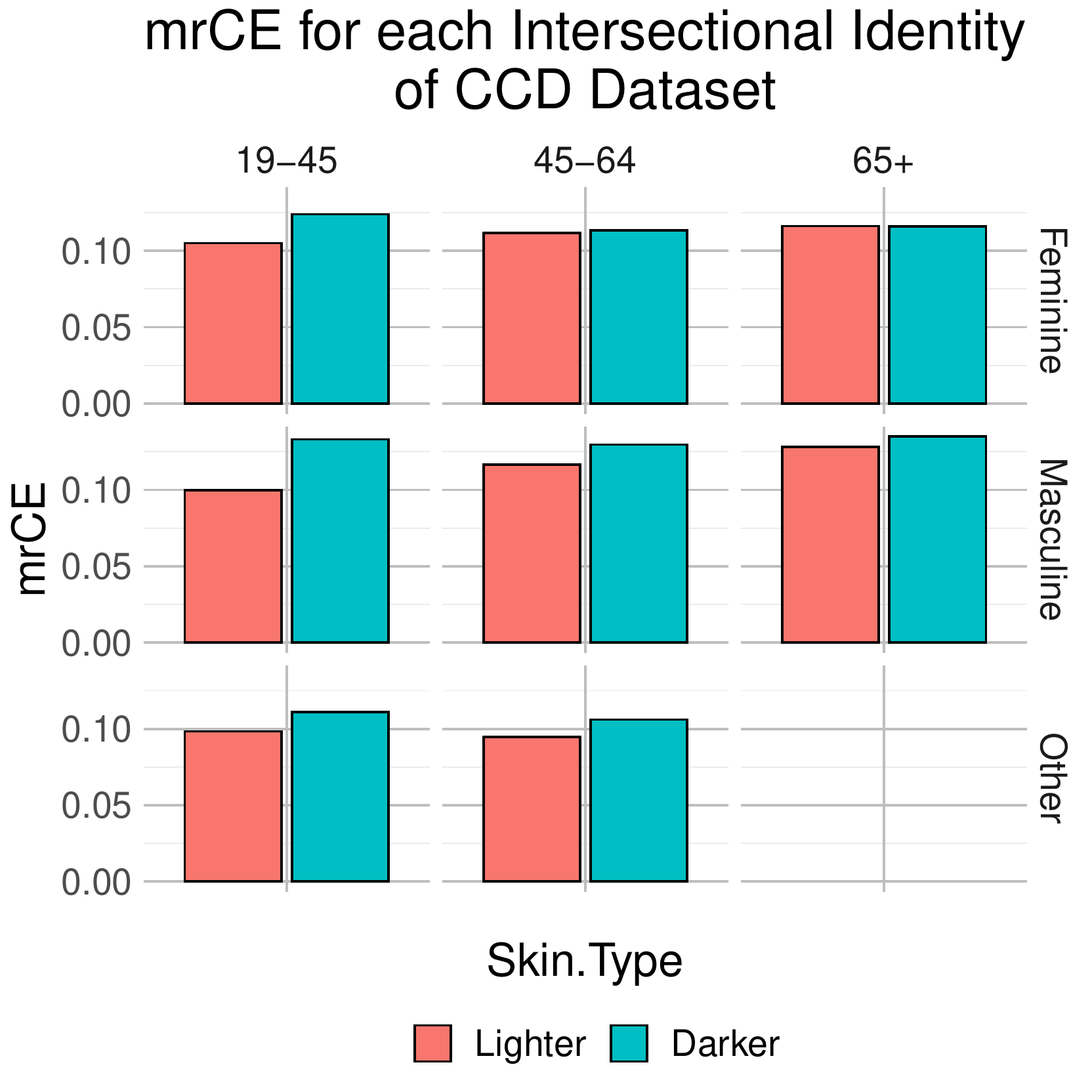}
\caption{In all intersectional identities, darker skinned individuals are less robust than those who are lighter skinned.}%
\label{fig:comp_skin}%
\end{minipage}%
\qquad
\begin{minipage}{.3\textwidth}%
  \centering
  \includegraphics[width=\textwidth]{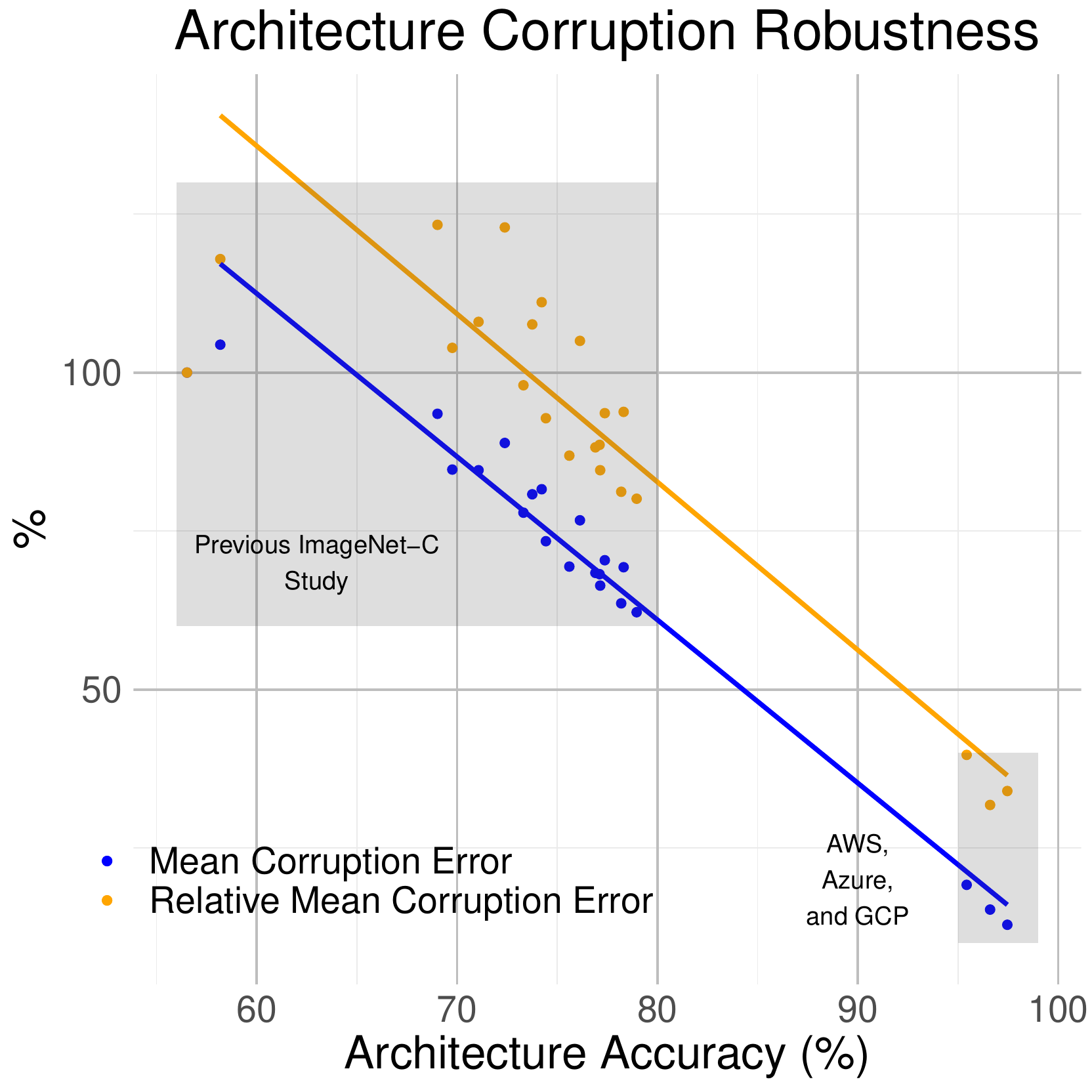}
\caption{ Recreation of Figure 3 from \citet{hendrycks2019benchmarking} with contemporary results and the addition of our findings.}%
\label{fig:imagenetc_comp}%
\end{minipage}%
\end{figure}

Across all datasets except UTKFace, we find that feminine presenting individuals have lower errors than masculine presenting individuals. See Figure~\ref{fig:comp_gender}. On Adience, feminine individuals have 18.8\% \mrCE{} whereas masculine have 19.8\%. On CCD, the $\mrCE{}$s are 8.9\% and 9.6\% respectively. On the MIAP dataset, the $\mrCE{}$ values are 13.7\% and 15.4\% respectively. On the UTKFace, both gender presentations have around 9.0\% $\mrCE{}$ (non statistically significant difference).

Stepping outside the gender binary, we have two insights into this from these data. In the CCD dataset, the subjects were asked to self-identify their gender. Two individuals selected Other and 62 others did not provide a response. Those two who chose outside the gender binary have a $\mrCE{}$ of 4.9\%. When we include those individuals without gender labels, their $\mrCE{}$ is 8.8\% and not significantly different from the feminine presenting individuals. 

The other insight comes from the MIAP dataset where subjects were rated on their perceived gender presentation by crowdworkers; options were ``Predominantly Feminine", ``Predominantly Masculine", and "Unknown". For those ``Unknown", the overall $\mrCE{}$ is 19.3\%. The creators of the dataset automatically set the gender presenation of those with an age presentation of ``Young" to be ``Unknown". The $\mrCE{}$ of those annotations which are not ``Young" and have an ``Unknown" gender presentation raises to 19.9\%. One factor that might contribute to this phenomenon is that individuals with an ``Unknown'' gender presentation might have faces that are occluded or are small in the image. Further work should be done to explore the causes of this discrepancy. 

\subsection{Dark skinned subjects have more errors across age and gender identities}

\begin{figure}
\centering
  \includegraphics[width=.39\textwidth, height=2.2in]{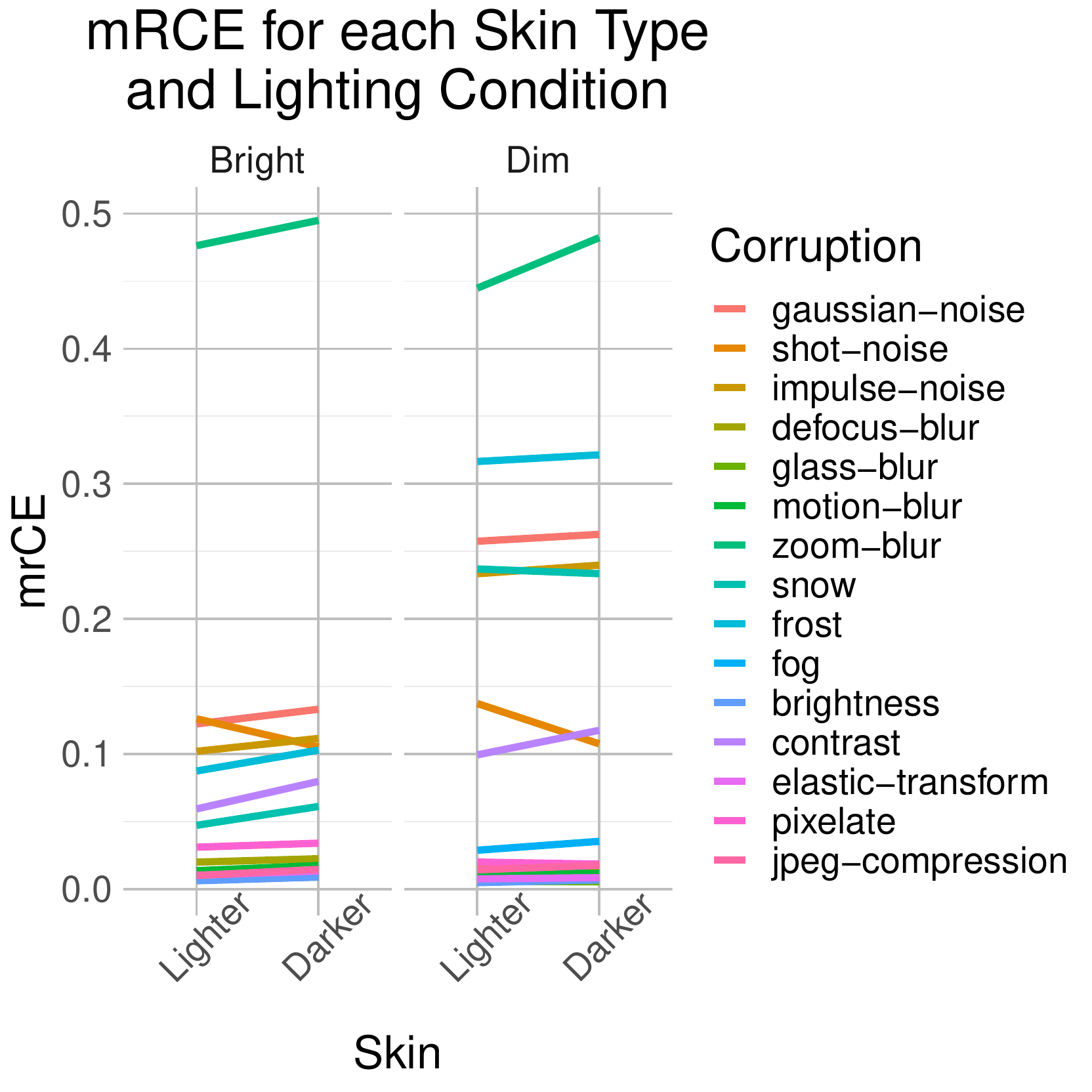}
  \includegraphics[width=.6\textwidth, height=2.2in]{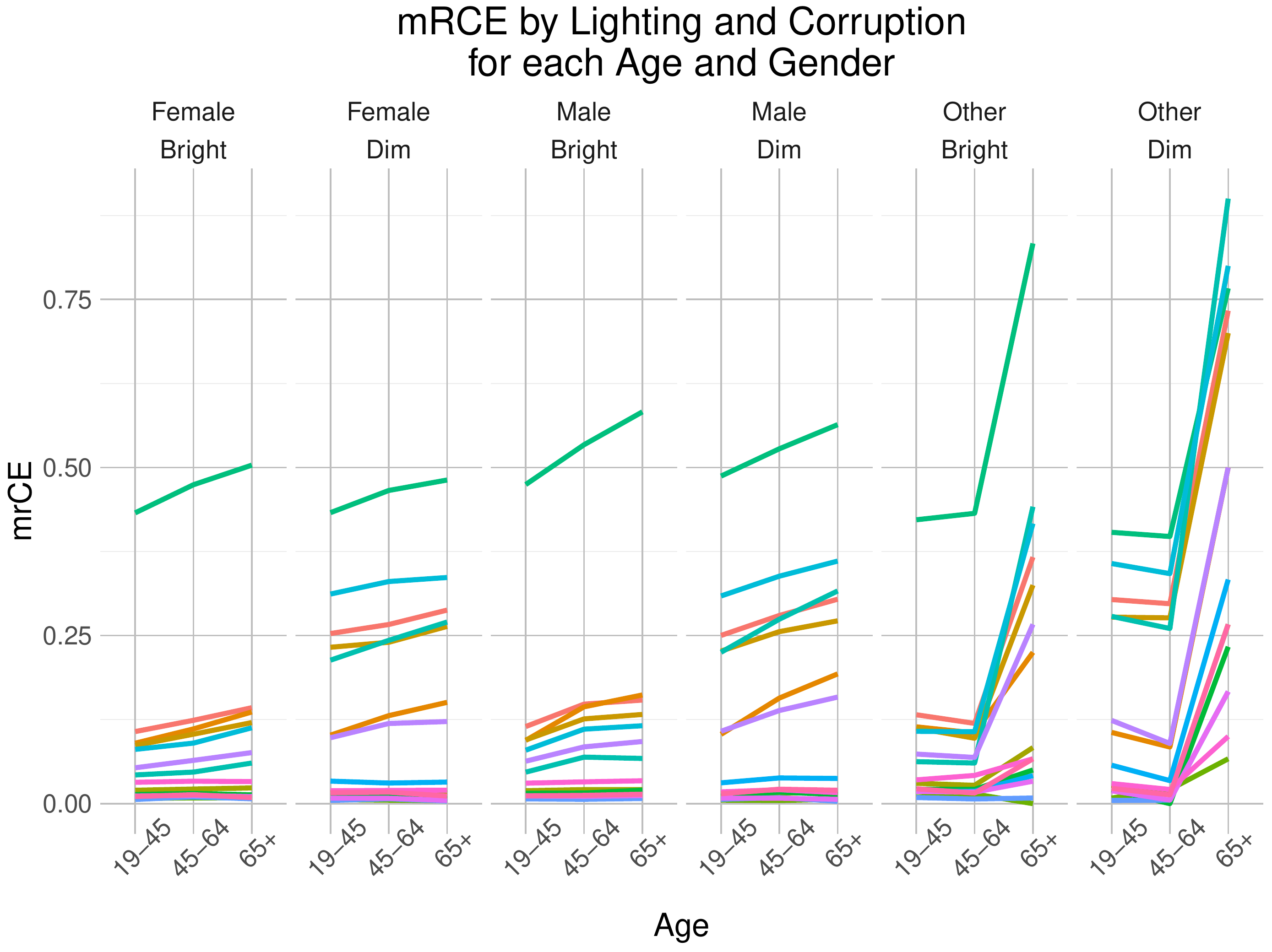}
\caption{(Left) \mrCE{} is plotted for each corruption by the intersection of lighting condition and skin type. (Right) the same is plotted by the intersection of age, gender, and lighting. Observe that for both skin types, all genders, and all ages, the dimly lit environment increases the error rates.}
\label{fig:comp_lighting}
\end{figure}

We analyze data from the CCD dataset which has ratings for each subject on the Fitzpatrick scale. As is customary in analyzing these ratings, we split the six Fitzpatrick values into two: Lighter (for ratings I-III) and Darker for ratings (IV-VI). The main intersectional results are reported in Figure~\ref{fig:comp_skin}.

The overall $\mrCE{}$ for lighter and darker skin types are 8.5\% and 9.7\% respectively, a 15\% increase for the darker skin type. We also see a similar trend in the intersectional identities available in the CCD metadata (age, gender, and skin type). We see that in every identity (except for 45-64 year old and Feminine) the darker skin type has statistically significant higher error rates. This difference is particularly stark in 19-45 year old, masculine subjects. We see a 35\% increase in errors for the darker skin type subjects in this identity compared to those with lighter skin types. For every 20 errors on a light skinned, masculine presenting individual between 18 and 45, there are 27 errors for dark skinned individuals of the same category.  

\subsection{Dim lighting conditions has the most severe impact on errors}

Using lighting condition information from the CCD dataset, we observe the $\mrCE{}$ is substantially higher in dimly lit environments: 12.5\% compared to 7.8\% in bright environments. See Figure~\ref{fig:comp_lighting}.

Across the board, we generally see that the disparity in demographic groups decreases between bright and dimly lit environments. For example, the odds ratio between dark and light skinned subjects is 1.09 for bright environments, but decreases to 1.03 for dim environments. This is true for age groups (e.g., odds ratios 1.150 (bright) vs 1.078 (dim) for 45-64 compared to 19-45; 1.126 (bright) vs 1.060 (dim) for Males compared to Females). This is not true for individuals with gender identities as Other or omitted -- the disparity increases (1.053 (bright) vs 1.145 (dim) with Females as the reference).

In Figure~\ref{fig:comp_lighting} we observe the lighting differences for different intersectional identities across corruptions. We continue to see zoom blur as the most challenging corruption. Interestingly, the noise and some weather corruptions have a large increase in their errors in dimly lit environments across intersectional identities whereas many of the other corruptions do not.

\subsection{Older subjects have higher gender error disparities}

\begin{figure}
    \centering
    \includegraphics[width=\textwidth]{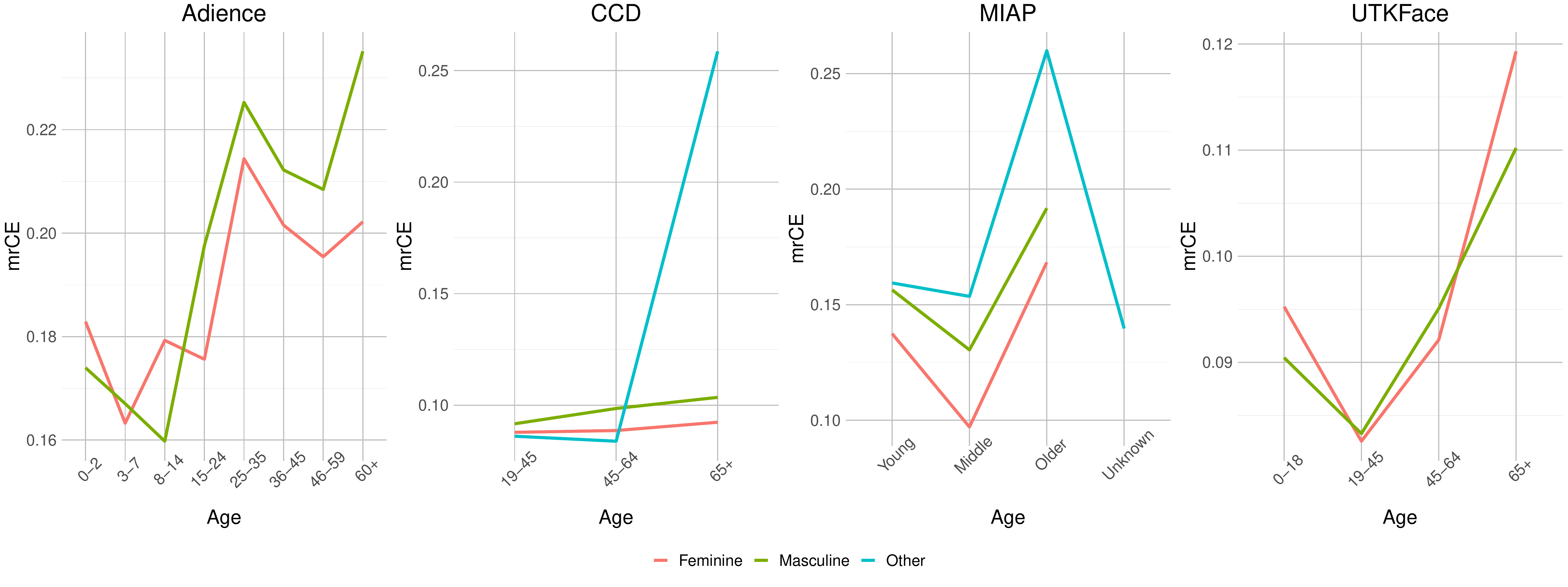}
    \caption{For each dataset, the \mrCE{} is plotted across age groups. Each gender is represented and indicates how gender disparities change across the age groups.}
    \label{fig:comp_age_gender}
\end{figure}

We plot in Figure~\ref{fig:comp_age_gender} the \mrCE{} for each dataset across age with each gender group plotted separately. From this, we can note that on the CCD and MIAP dataset, the masculine presenting group is always less robust than the feminine. On the CCD dataset, the disparity between the two groups increases as the age increases (odds ratio of 1.040 for 19-45 raises to 1.138 for 65+). On the MIAP dataset, the odds ratio is greatest between masculine and feminine for the middle age group (1.395). The disparities between the ages also increases from feminine to masculine to unknown gender identities. 

On the Adience and UTKFace datasets, we see that the feminine presenting individuals sometimes have higher error rates than masculine presenting subjects. Notably, the most disparate errors in genders on these datasets occurs at the oldest categories, following the trend from the other datasets.

\section{Gender and Age Estimation Analysis}\label{sec:age_gender_prediction}

We briefly overview results from evaluating AWS's age and gender estimation commercial systems. Further analysis can be found in Appendices~\ref{sec:gender_pred} and \ref{sec:age_est}.

\subsection{Gender estimation is at least twice as susceptible to corruptions as face detection}

The use of automated gender estimates in ML is a controversial topic. Trans and gender queer individuals are often ignored in ML research, though there is a growing body of research that aims to use these technologies in an assistive way as well \citep[e.g.,][]{ahmed2019bridging,chong2021exploring}.
To evaluate gender estimation, we only use CCD as the subjects of these photos voluntarily identified their gender. We omit from the analysis any individual who either did not choose to give their gender or falls outside the gender binary because AWS only estimates Male and Female.

AWS misgenders 9.1\% of the clean images but 21.6\% of the corrupted images. Every corruption performs worse on gender estimation than \mrCE{}. Two corruptions (elastic transform and glass blur) do not have statistically different errors from the clean images. All the others do, with the most significant being zoom blur, Gaussian noise, impulse noise, snow, frost, shot noise, and contrast. Zoom blur's probability of error is 61\% and Gaussian noise is 32\%. This compares to \mrCE{} values of 43\% and 29\% respectively. See Appendix~\ref{sec:gender_pred} for further analysis. 

\subsection{Corrupted images error in their age predictions by 40\% more than clean images}

To estimate Age, AWS returns an upper and lower age estimation. Following their own guidelines on face detection~\citep{AWSGuidelines}, we use the mid-point of these numbers as a approximate estimate. On average, the estimation is 8.3 years away from the actual age of the subject for corrupted data, this compares to 5.9 years away for clean data. See Appendix~\ref{sec:age_est} for further analysis.

\section{Conclusion}

This benchmark has evaluated three leading commercial facial detection and analysis systems for their robustness against common natural noise corruptions. Using the 15 ImageNet-C corruptions, we measured the relative mean corruption error as measured by comparing the number of faces detected in a clean and corrupted image. We used four academic datasets which included demographic detail. 

We observed through our analysis that there are significant demographic disparities in the likelihood of error on corrupted data. We found that older individuals, masculine presenting individuals, those with darker skin types, or in photos with dim ambient light all have higher errors ranging from 20-60\%. We also investigated questions of intersectional identities finding that darker males have the highest corruption errors. As for age and gender estimation, corruptions have a significant and sizeable impact on the system's performance; gender estimation is more than twice as bad on corrupted images as it is on clean images; age estimation is 40\% worse on corrupted images. 

Future work could explore other metrics for evaluating face detection systems when ground truth bounding boxes are not present. While we considered the length of response on clean images to be ground truth, it could be viable to treat the clean image's bounding boxes as ground truth and measure deviations therefrom when considering questions of robustness. Of course, this would require a transition to detection-based metrics like precision, recall, and $F$-measure. 

We do not explore questions of causation in this benchmark. We do not have enough different datasets or commercial systems to probe this question through regressions or mixed effects modeling. We do note that there is work that examines causation questions with such methods like that of \citet{best2017longitudinal} and \citet{cook2019demographic}. With additional data and under similar benchmarking protocols, one could start to examine this question. However, the black-box nature of commercial systems presents unique challenges to this endeavor. 

\clearpage
\begin{ack}

This research was supported in part by NSF CAREER Award IIS-1846237, NSF D-ISN Award \#2039862, NSF Award CCF-1852352, NIH R01 Award NLM-013039-01, NIST MSE Award \#20126334, DARPA GARD \#HR00112020007, and DoD WHS Award \#HQ003420F0035.  We thank Candice Schumann for answering questions related to the MIAP dataset, as well as Aurelia Augusta, Brian Brubach, Valeria Cherepanova, Vedant Nanda, Aviva Prins, Liz O'Sullivan, Neehar Peri, Candice Schumann for advice and feedback.
\end{ack}

\bibliographystyle{abbrvnat}
\bibliography{refs.bib}

\appendix

\section{Evaluation Information}

\subsection{Image Counts}\label{sec:image_counts}

For each dataset, we selected no more than \num{1500} images from any intersectional group. The final tallies of how many images from each group can be found in Tables~\ref{tbl:image_count_breakdown_adience}, \ref{tbl:image_count_breakdown_ccd}, \ref{tbl:image_count_breakdown_miap}, and \ref{tbl:image_count_breakdown_utk}.

{\color{black}
\subsection{Corruption information}

We evaluate 15 corruptions from \citet{hendrycks2019benchmarking}: Gaussian noise, shot noise, impulse noise, defocus blur, glass blur, motion blur, zoom blur, snow, frost, fog, brightness, contrast, elastic transforms, pixelation, and jpeg compressions. Each corruption is described in the \citet{hendrycks2019benchmarking} paper as follows:

The first corruption type is Gaussian noise. This corruption can appear
in low-lighting conditions. Shot noise, also called Poisson noise, is electronic noise caused by the
discrete nature of light itself. Impulse noise is a color analogue of salt-and-pepper noise and can be
caused by bit errors. Defocus blur occurs when an image is out of focus. Frosted Glass Blur appears
with “frosted glass” windows or panels. Motion blur appears when a camera is moving quickly. Zoom
blur occurs when a camera moves toward an object rapidly. Snow is a visually obstructive form of
precipitation. Frost forms when lenses or windows are coated with ice crystals. Fog shrouds objects
and is rendered with the diamond-square algorithm. Brightness varies with daylight intensity. Contrast
can be high or low depending on lighting conditions and the photographed object’s color. Elastic
transformations stretch or contract small image regions. Pixelation occurs when upsampling a lowresolution image. JPEG is a lossy image compression format which introduces compression artifacts.

The specific parameters for each corruption can be found in the project's github at the corruptions file: \url{https://github.com/dooleys/Robustness-Disparities-in-Commercial-Face-Detection/blob/main/code/imagenet_c_big/corruptions.py}.
}

\begin{table}
\caption{\footnotesize Adience Dataset Counts}
\label{tbl:image_count_breakdown_adience}
\centering
\begin{tabular}[t]{llr}
\toprule
     &      &     Count \\
Age & Gender &       \\
\midrule
0-2 & Female &   684 \\
     & Male &   716 \\
3-7 & Female &  1232 \\
     & Male &   925 \\
8-14 & Female &  1353 \\
     & Male &   933 \\
15-24 & Female &  1047 \\
     & Male &   742 \\
25-35 & Female &  1500 \\
     & Male &  1500 \\
36-45 & Female &  1078 \\
     & Male &  1412 \\
46-59 & Female &   436 \\
     & Male &   466 \\
60+ & Female &   428 \\
     & Male &   467 \\
\bottomrule
\end{tabular}
\end{table}

\begin{table}
\centering
\caption{\footnotesize CCD Dataset Counts}
\label{tbl:image_count_breakdown_ccd}
\begin{tabular}{llllr}
\toprule
    &       &       &       &     Count \\
Lighting & Gender & Skin & Age &       \\
\midrule
Bright & Female & Dark & 19-45 &  1500 \\
    &       &       & 45-64 &  1500 \\
    &       &       & 65+ &   547 \\
    &       & Light & 19-45 &  1500 \\
    &       &       & 45-64 &  1500 \\
    &       &       & 65+ &   653 \\
    & Male & Dark & 19-45 &  1500 \\
    &       &       & 45-64 &  1500 \\
    &       &       & 65+ &   384 \\
    &       & Light & 19-45 &  1500 \\
    &       &       & 45-64 &  1500 \\
    &       &       & 65+ &   695 \\
    & Other & Dark & 19-45 &   368 \\
    &       &       & 45-64 &   168 \\
    &       &       & 65+ &    12 \\
    &       & Light & 19-45 &   244 \\
    &       &       & 45-64 &    49 \\
Dim & Female & Dark & 19-45 &  1500 \\
    &       &       & 45-64 &   670 \\
    &       &       & 65+ &   100 \\
    &       & Light & 19-45 &   642 \\
    &       &       & 45-64 &   314 \\
    &       &       & 65+ &   131 \\
    & Male & Dark & 19-45 &  1500 \\
    &       &       & 45-64 &   387 \\
    &       &       & 65+ &    48 \\
    &       & Light & 19-45 &   485 \\
    &       &       & 45-64 &   299 \\
    &       &       & 65+ &   123 \\
    & Other & Dark & 19-45 &    57 \\
    &       &       & 45-64 &    26 \\
    &       &       & 65+ &     3 \\
    &       & Light & 19-45 &    27 \\
    &       &       & 45-64 &    12 \\
\bottomrule
\end{tabular}
\end{table}

\bigskip 

\begin{table}
\caption{\footnotesize MIAP Dataset Counts}
\label{tbl:image_count_breakdown_miap}
\centering
\begin{tabular}{llr}
\toprule
      &         &     Count \\
AgePresentation & GenderPresentation &       \\
\midrule
Young & Unknown &  1500 \\
Middle & Predominantly Feminine &  1500 \\
      & Predominantly Masculine &  1500 \\
      & Unknown &   561 \\
Older & Predominantly Feminine &   209 \\
      & Predominantly Masculine &   748 \\
      & Unknown &    24 \\
Unknown & Predominantly Feminine &   250 \\
      & Predominantly Masculine &   402 \\
      & Unknown &  1500 \\
\bottomrule
\end{tabular}
\end{table}

\begin{table}[!htbp]
\centering
\caption{\footnotesize UTKFace Dataset Counts}
\label{tbl:image_count_breakdown_utk}
\begin{tabular}{lllr}
\toprule
    &      &       &     Count \\
Age & Gender & Race &       \\
\midrule
0-18 & Female & Asian &   555 \\
    &      & Black &   161 \\
    &      & Indian &   350 \\
    &      & Others &   338 \\
    &      & White &   987 \\
    & Male & Asian &   586 \\
    &      & Black &   129 \\
    &      & Indian &   277 \\
    &      & Others &   189 \\
    &      & White &   955 \\
19-45 & Female & Asian &  1273 \\
    &      & Black &  1500 \\
    &      & Indian &  1203 \\
    &      & Others &   575 \\
    &      & White &  1500 \\
    & Male & Asian &   730 \\
    &      & Black &  1499 \\
    &      & Indian &  1264 \\
    &      & Others &   477 \\
    &      & White &  1500 \\
45-64 & Female & Asian &    39 \\
    &      & Black &   206 \\
    &      & Indian &   146 \\
    &      & Others &    22 \\
    &      & White &   802 \\
    & Male & Asian &   180 \\
    &      & Black &   401 \\
    &      & Indian &   653 \\
    &      & Others &    97 \\
    &      & White &  1500 \\
65+ & Female & Asian &    75 \\
    &      & Black &    78 \\
    &      & Indian &    43 \\
    &      & Others &    10 \\
    &      & White &   712 \\
    & Male & Asian &   148 \\
    &      & Black &   166 \\
    &      & Indian &    91 \\
    &      & Others &     5 \\
    &      & White &   682 \\
\bottomrule
\end{tabular}
\end{table}

\section{Metric Discussion}\label{sec:metric_example}

Our use of relative error is slightly adapted from ImageNet-C insomuch that in that paper, they were measuring top-1 error of classification systems. However, the concept is identical. Consequently, it is linguistically best for our measure to be \emph{mean relative} corruption error, whereas ImageNet-C reports the \emph{mean relative} corruption error. This symmantic difference is attributed to when we take our average versus when the ImageNet-C protocol does.

{\color{black}
\subsection{Response Example}\label{sec:response:example}
Our main metric, \mrCE{} relies on the computing the number of detected faces, i.e.,length of a response, $l_r$, from an API service. We explicitly give an example of this here. 

An example response from an API service is below has one face detected:
\begin{lstlisting}
[
  {
    "face_rectangle": {
      "width": 601,
      "height": 859,
      "left": 222,
      "top": 218
    }
  }
]
\end{lstlisting}

An example response from an API service is below has two faces detected:
\begin{lstlisting}
[
  {
    "face_rectangle": {
      "width": 601,
      "height": 859,
      "left": 222,
      "top": 218
    }
  },
  {
    "face_rectangle": {
      "width": 93,
      "height": 120,
      "left": 10,
      "top": 39
    }
  }  
]
\end{lstlisting}

}

\subsection{Metric Example}
Let us consider an example. Assume we were testing the question of whether there is a difference in error rates for different eye colors: (Grey, Hazel, and Brown). Across all the corrupted data, we might see that $\rCE{}$ is 0.12, 0.21, and 0.23 for Grey, Hazel, and Brown respectively. Recall that the odds of an event with likelihood $p$ is reported as $p/(1-p)$. So the odds of error for each eye color is 0.14, 0.27, and 0.30 respectively. Our logistic regression would be written as $\texttt{rCE} = \beta_0 + \beta_1\texttt{Hazel}+\beta_2\texttt{Brown}$, with \texttt{Hazel} and \texttt{Brown} being indicator variables. After fitting our regression, we might see that the estimated \emph{odds} coefficients for the intercept is 0.14, for the variable \texttt{Hazel} is 1.93, and for the variable Brown is 2.14; and all the coefficients are significant. This makes sense because the odds of Grey is 0.14, the odds ratio between Hazel and Grey is 0.27/0.14 = 1.93, and the odds ratio between Brown and Grey is 0.30/0.14 = 2.14. The significance tells us that the odds (or probability) or error for Grey eyes is significantly different from the odds (or probability) of error for Brown and Hazel eyes. In this example, we can conclude that the odds of error is 2.14 higher for Brown eyes compared to Grey. Put another way, for every 1 error for a Grey eyed person, there would be roughly 2 errors for a Hazel or Brown person. More so, the odds of error for Brown eyes is 114\% higher than the odds of error for for Grey eyes.

\section{API Parameteres}\label{sec:api}

For the AWS DetectFaces API,\footnote{\url{https://docs.aws.amazon.com/rekognition/latest/dg/API_DetectFaces.html}} we selected to have all facial attributes returned. This includes age and gender estimates. We evaluate the performance of these estimates in Section \ref{sec:age_gender_prediction}. The Azure Face API\footnote{\url{https://westus.dev.cognitive.microsoft.com/docs/services/563879b61984550e40cbbe8d/operations/563879b61984550f30395236}} allows the user to select one of three detection models. We chose model \texttt{detection\_03} as it was their most recently released model (February 2021) and was described to have the highest performance on small, side, and blurry faces, since it aligns with our benchmark intention. This model does not return age or gender estimates (though model \texttt{detection\_01} does).

\section{Benchmarks Costs}\label{sec:costs}

A total breakdown of costs for this benchmark can be found in Table~\ref{tbl:costs}.

\begin{table}[!htbp]
\centering
\caption{\footnotesize Total Costs of Benchmark}
\label{tbl:costs}
\begin{tabular}{lr}
\toprule
Category & Cost \\
\midrule
Azure Face Service & \$4,270.58 \\
AWS Rekognition & \$4,270.66\\
Google Cloud Platform & \$7,230.47\\
S3 & \$1,003.83\\
EC2 & \$475.77\\
Tax & \$256.24\\
\midrule\\
Total & \$17,507.55\\
\bottomrule
\end{tabular}
\end{table}

\section{Statistical Significance Regressions for \rCE}

\subsection{Main Tables}

Regression of \mrCE{} for all variables for the all the datasets with unified demographic variables can be found in Table~\ref{tbl:comp_full}.

\begin{table}[!htbp] \centering 
  \caption{Main regressions (odds ratio) for the all datasets with unifed demographic variables.} 
  \label{tbl:comp_full} 
\resizebox{!}{11cm}{
\begin{tabular}{@{\extracolsep{5pt}}lcccc} 
\\[-1.8ex]\hline 
\hline \\[-1.8ex] 
 & \multicolumn{4}{c}{\textit{Dependent variable:}} \\ 
\cline{2-5} 
\\[-1.8ex] & \multicolumn{4}{c}{rCE} \\ 
 & Adience & CCD & MIAP & UTKFace \\ 
\\[-1.8ex] & (1) & (2) & (3) & (4)\\
\end{tabular} }
\end{table}

Regression of \mrCE{} for all variables for the Adience dataset can be found in Table~\ref{tbl:comp_full_adience}.

\begin{table}[!htbp] \centering 
  \caption{Full regression (odds ratio) for the Adience dataset.} 
  \label{tbl:comp_full_adience} 
\resizebox{!}{11cm}{
\begin{tabular}{@{\extracolsep{5pt}}lcccc} 
\\[-1.8ex]\hline 
\hline \\[-1.8ex] 
 & \multicolumn{4}{c}{\textit{Dependent variable:}} \\ 
\cline{2-5} 
\\[-1.8ex] & \multicolumn{4}{c}{rCE} \\ 
 & All Data & Just AWS & Just Azure & Just GCP \\ 
\\[-1.8ex] & (1) & (2) & (3) & (4)\\
\end{tabular} }
\end{table}  

Regression of \mrCE{} for all variables for the CCD dataset can be found in Table~\ref{tbl:comp_full_ccd}.

\begin{table}[!htbp] \centering 
  \caption{Full regression (odds ratio) for the CCD dataset.} 
  \label{tbl:comp_full_ccd} 
\resizebox{!}{11cm}{
\begin{tabular}{@{\extracolsep{5pt}}lcccc} 
\\[-1.8ex]\hline 
\hline \\[-1.8ex] 
 & \multicolumn{4}{c}{\textit{Dependent variable:}} \\ 
\cline{2-5} 
\\[-1.8ex] & \multicolumn{4}{c}{rCE} \\ 
 & All Data & Just AWS & Just Azure & Just GCP \\ 
\\[-1.8ex] & (1) & (2) & (3) & (4)\\
\end{tabular} }
\end{table}  

Regression of \mrCE{} for all variables for the MIAP dataset can be found in Table~\ref{tbl:comp_full_miap}.

\begin{table}[!htbp] \centering 
  \caption{Full regression (odds ratio) for the MIAP dataset.} 
  \label{tbl:comp_full_miap} 
\resizebox{!}{11cm}{
\begin{tabular}{@{\extracolsep{5pt}}lcccc} 
\\[-1.8ex]\hline 
\hline \\[-1.8ex] 
 & \multicolumn{4}{c}{\textit{Dependent variable:}} \\ 
\cline{2-5} 
\\[-1.8ex] & \multicolumn{4}{c}{rCE} \\ 
 & All Data & Just AWS & Just Azure & Just GCP \\ 
\\[-1.8ex] & (1) & (2) & (3) & (4)\\
\end{tabular} }
\end{table}

Regression of \mrCE{} for all variables for the UTKFace dataset can be found in Table~\ref{tbl:comp_full_utk}.

\begin{table}[!htbp] \centering 
  \caption{Full regression (odds ratio) for the UTKFace dataset.} 
  \label{tbl:comp_full_utk} 
\resizebox{!}{11cm}{
\begin{tabular}{@{\extracolsep{5pt}}lcccc} 
\\[-1.8ex]\hline 
\hline \\[-1.8ex] 
 & \multicolumn{4}{c}{\textit{Dependent variable:}} \\ 
\cline{2-5} 
\\[-1.8ex] & \multicolumn{4}{c}{rCE} \\ 
 & All Data & Just AWS & Just Azure & Just GCP \\ 
\\[-1.8ex] & (1) & (2) & (3) & (4)\\
\end{tabular} }
\end{table}

\subsubsection{Adience Demographic Interactions}\label{sec:comp_int_adience}
Regression of \mrCE{} for the interaction of Age and Gender for the Adience dataset can be found in Table~\ref{tbl:comp_int_adience_1}.

\begin{table}[!htbp] \centering 
  \caption{Interaction (odds ratio) of Age and Gender for the Adience dataset.} 
  \label{tbl:comp_int_adience_1} 
\resizebox{\textwidth}{!}{
\begin{tabular}{@{\extracolsep{5pt}}lccccccccc} 
\\[-1.8ex]\hline 
\hline \\[-1.8ex] 
 & \multicolumn{9}{c}{\textit{Dependent variable:}} \\ 
\cline{2-10} 
\\[-1.8ex] & \multicolumn{9}{c}{rCE} \\ 
 & Interaction & Just 0-2 & Just 3-7 & Just 8-14 & Just 15-24 & Just 25-35 & Just 36-45 & Just 46-59 & Just 60+ \\ 
\\[-1.8ex] & (1) & (2) & (3) & (4) & (5) & (6) & (7) & (8) & (9)\\
\end{tabular} }
\end{table}

\subsubsection{CCD Demographic Interactions}\label{sec:comp_int_ccd}
Regression of \mrCE{} for the interaction of Lighting and Skin Type for the CCD dataset can be found in Table~\ref{tbl:comp_int_ccd_1}.

\begin{table}[!htbp] \centering 
  \caption{Interaction (odds ratio) of Lighting and Skin Type for the CCD dataset.} 
  \label{tbl:comp_int_ccd_1} 
\begin{tabular}{@{\extracolsep{5pt}}lccc} 
\\[-1.8ex]\hline 
\hline \\[-1.8ex] 
 & \multicolumn{3}{c}{\textit{Dependent variable:}} \\ 
\cline{2-4} 
\\[-1.8ex] & \multicolumn{3}{c}{rCE} \\ 
 & Interaction & Just dark Lighting & Just light Lighting \\ 
\\[-1.8ex] & (1) & (2) & (3)\\ 
\hline \\[-1.8ex] 
 lightingDark & 1.804 &  &  \\ 
  & t = 115.964$^{***}$ &  &  \\ 
  & & & \\ 
 FitzDark Fitz & 1.083 & 1.027 & 1.083 \\ 
  & t = 20.524$^{***}$ & t = 5.111$^{***}$ & t = 20.524$^{***}$ \\ 
  & & & \\ 
 lightingDark:FitzDark Fitz & 0.948 &  &  \\ 
  & t = $-$8.234$^{***}$ &  &  \\ 
  & & & \\ 
 Constant & 0.090 & 0.162 & 0.090 \\ 
  & t = $-$868.977$^{***}$ & t = $-$426.198$^{***}$ & t = $-$868.977$^{***}$ \\ 
  & & & \\ 
\hline \\[-1.8ex] 
Observations & 4,824,900 & 1,422,900 & 3,402,000 \\ 
Log Likelihood & $-$1,575,840.000 & $-$581,078.700 & $-$994,761.100 \\ 
Akaike Inf. Crit. & 3,151,688.000 & 1,162,161.000 & 1,989,526.000 \\ 
\hline 
\hline \\[-1.8ex] 
\textit{Note:}  & \multicolumn{3}{r}{$^{*}$p$<$0.1; $^{**}$p$<$0.05; $^{***}$p$<$0.01} \\ 
\end{tabular} 
\end{table}  

Regression of \mrCE{} for the interaction of Lighting and Age for the CCD dataset can be found in Table~\ref{tbl:comp_int_ccd_2}.

\begin{table}[!htbp] \centering 
  \caption{Interaction (odds ratio) of Lighting and Age for the CCD dataset.} 
  \label{tbl:comp_int_ccd_2} 
\begin{tabular}{@{\extracolsep{5pt}}lccc} 
\\[-1.8ex]\hline 
\hline \\[-1.8ex] 
 & \multicolumn{3}{c}{\textit{Dependent variable:}} \\ 
\cline{2-4} 
\\[-1.8ex] & \multicolumn{3}{c}{rCE} \\ 
 & Interaction & Just dark Lighting & Just light Lighting \\ 
\\[-1.8ex] & (1) & (2) & (3)\\ 
\hline \\[-1.8ex] 
 lightingDark & 1.887 &  &  \\ 
  & t = 148.921$^{***}$ &  &  \\ 
  & & & \\ 
 Age45-64 & 1.150 & 1.078 & 1.150 \\ 
  & t = 32.852$^{***}$ & t = 13.765$^{***}$ & t = 32.852$^{***}$ \\ 
  & & & \\ 
 Age65+ & 1.284 & 1.178 & 1.284 \\ 
  & t = 44.789$^{***}$ & t = 17.206$^{***}$ & t = 44.789$^{***}$ \\ 
  & & & \\ 
 lightingDark:Age45-64 & 0.937 &  &  \\ 
  & t = $-$9.414$^{***}$ &  &  \\ 
  & & & \\ 
 lightingDark:Age65+ & 0.918 &  &  \\ 
  & t = $-$7.783$^{***}$ &  &  \\ 
  & & & \\ 
 Constant & 0.085 & 0.160 & 0.085 \\ 
  & t = $-$807.913$^{***}$ & t = $-$614.951$^{***}$ & t = $-$807.913$^{***}$ \\ 
  & & & \\ 
\hline \\[-1.8ex] 
Observations & 4,824,900 & 1,422,900 & 3,402,000 \\ 
Log Likelihood & $-$1,574,722.000 & $-$580,886.600 & $-$993,835.600 \\ 
Akaike Inf. Crit. & 3,149,456.000 & 1,161,779.000 & 1,987,677.000 \\ 
\hline 
\hline \\[-1.8ex] 
\textit{Note:}  & \multicolumn{3}{r}{$^{*}$p$<$0.1; $^{**}$p$<$0.05; $^{***}$p$<$0.01} \\ 
\end{tabular} 
\end{table}  

Regression of \mrCE{} for the interaction of Lighting and Age (as a numeric variable) for the CCD dataset can be found in Table~\ref{tbl:comp_int_ccd_2.n}.

\begin{table}[!htbp] \centering 
  \caption{Interaction (odds ratio) of Lighting and Age (as a numeric variable) for the CCD dataset.} 
  \label{tbl:comp_int_ccd_2.n} 
\begin{tabular}{@{\extracolsep{5pt}}lc} 
\\[-1.8ex]\hline 
\hline \\[-1.8ex] 
 & \multicolumn{1}{c}{\textit{Dependent variable:}} \\ 
\cline{2-2} 
\\[-1.8ex] & rCE \\ 
\hline \\[-1.8ex] 
 lightingDark & 1.952 \\ 
  & t = 80.718$^{***}$ \\ 
  & \\ 
 Age\_Numeric & 1.005 \\ 
  & t = 48.335$^{***}$ \\ 
  & \\ 
 lightingDark:Age\_Numeric & 0.998 \\ 
  & t = $-$9.639$^{***}$ \\ 
  & \\ 
 Constant & 0.075 \\ 
  & t = $-$504.962$^{***}$ \\ 
  & \\ 
\hline \\[-1.8ex] 
Observations & 4,824,900 \\ 
Log Likelihood & $-$1,574,663.000 \\ 
Akaike Inf. Crit. & 3,149,334.000 \\ 
\hline 
\hline \\[-1.8ex] 
\textit{Note:}  & \multicolumn{1}{r}{$^{*}$p$<$0.1; $^{**}$p$<$0.05; $^{***}$p$<$0.01} \\ 
\end{tabular} 
\end{table}  

Regression of \mrCE{} for the interaction of Lighting and Gender for the CCD dataset can be found in Table~\ref{tbl:comp_int_ccd_3}.

\begin{table}[!htbp] \centering 
  \caption{Interaction (odds ratio) of Lighting and Gender for the CCD dataset.} 
  \label{tbl:comp_int_ccd_3} 
\begin{tabular}{@{\extracolsep{5pt}}lccc} 
\\[-1.8ex]\hline 
\hline \\[-1.8ex] 
 & \multicolumn{3}{c}{\textit{Dependent variable:}} \\ 
\cline{2-4} 
\\[-1.8ex] & \multicolumn{3}{c}{rCE} \\ 
 & Interaction & Just dark Lighting & Just light Lighting \\ 
\\[-1.8ex] & (1) & (2) & (3)\\ 
\hline \\[-1.8ex] 
 lightingDark & 1.819 &  &  \\ 
  & t = 135.888$^{***}$ &  &  \\ 
  & & & \\ 
 GenderMale & 1.126 & 1.060 & 1.126 \\ 
  & t = 29.668$^{***}$ & t = 11.895$^{***}$ & t = 29.668$^{***}$ \\ 
  & & & \\ 
 GenderOther & 1.053 & 1.145 & 1.053 \\ 
  & t = 5.870$^{***}$ & t = 8.063$^{***}$ & t = 5.870$^{***}$ \\ 
  & & & \\ 
 lightingDark:GenderMale & 0.941 &  &  \\ 
  & t = $-$9.608$^{***}$ &  &  \\ 
  & & & \\ 
 lightingDark:GenderOther & 1.088 &  &  \\ 
  & t = 4.447$^{***}$ &  &  \\ 
  & & & \\ 
 Constant & 0.088 & 0.161 & 0.088 \\ 
  & t = $-$843.455$^{***}$ & t = $-$548.811$^{***}$ & t = $-$843.455$^{***}$ \\ 
  & & & \\ 
\hline \\[-1.8ex] 
Observations & 4,824,900 & 1,422,900 & 3,402,000 \\ 
Log Likelihood & $-$1,575,531.000 & $-$581,000.400 & $-$994,530.700 \\ 
Akaike Inf. Crit. & 3,151,074.000 & 1,162,007.000 & 1,989,067.000 \\ 
\hline 
\hline \\[-1.8ex] 
\textit{Note:}  & \multicolumn{3}{r}{$^{*}$p$<$0.1; $^{**}$p$<$0.05; $^{***}$p$<$0.01} \\ 
\end{tabular} 
\end{table}  

Regression of \mrCE{} for the interaction of Age and Gender for the CCD dataset can be found in Table~\ref{tbl:comp_int_ccd_4}.

\begin{table}[!htbp] \centering 
  \caption{Interaction (odds ratio) of Age and Gender for the CCD dataset.} 
  \label{tbl:comp_int_ccd_4} 
\resizebox{\textwidth}{!}{
\begin{tabular}{@{\extracolsep{5pt}}lcccc} 
\\[-1.8ex]\hline 
\hline \\[-1.8ex] 
 & \multicolumn{4}{c}{\textit{Dependent variable:}} \\ 
\cline{2-5} 
\\[-1.8ex] & \multicolumn{4}{c}{rCE} \\ 
 & Interaction & Just 19-45 & Just 45-64 & Just 65+ \\ 
\\[-1.8ex] & (1) & (2) & (3) & (4)\\ 
\hline \\[-1.8ex] 
 Age45-64 & 0.967 &  &  &  \\ 
  & t = $-$7.088$^{***}$ &  &  &  \\ 
  & & & & \\ 
 Age65+ & 1.005 &  &  &  \\ 
  & t = 0.682 &  &  &  \\ 
  & & & & \\ 
 GenderMale & 1.040 & 1.040 & 1.107 & 1.138 \\ 
  & t = 9.019$^{***}$ & t = 9.019$^{***}$ & t = 20.113$^{***}$ & t = 15.471$^{***}$ \\ 
  & & & & \\ 
 GenderOther & 0.907 & 0.907 & 0.922 & 3.074 \\ 
  & t = $-$10.490$^{***}$ & t = $-$10.490$^{***}$ & t = $-$5.404$^{***}$ & t = 28.170$^{***}$ \\ 
  & & & & \\ 
 Age45-64:GenderMale & 1.064 &  &  &  \\ 
  & t = 9.266$^{***}$ &  &  &  \\ 
  & & & & \\ 
 Age65+:GenderMale & 1.094 &  &  &  \\ 
  & t = 9.519$^{***}$ &  &  &  \\ 
  & & & & \\ 
 Age45-64:GenderOther & 1.016 &  &  &  \\ 
  & t = 0.929 &  &  &  \\ 
  & & & & \\ 
 Age65+:GenderOther & 3.388 &  &  &  \\ 
  & t = 29.814$^{***}$ &  &  &  \\ 
  & & & & \\ 
 Constant & 0.112 & 0.112 & 0.108 & 0.112 \\ 
  & t = $-$708.886$^{***}$ & t = $-$708.886$^{***}$ & t = $-$625.073$^{***}$ & t = $-$373.867$^{***}$ \\ 
  & & & & \\ 
\hline \\[-1.8ex] 
Observations & 4,824,900 & 2,435,175 & 1,783,125 & 606,600 \\ 
Log Likelihood & $-$1,591,531.000 & $-$799,698.200 & $-$585,369.000 & $-$206,464.100 \\ 
Akaike Inf. Crit. & 3,183,080.000 & 1,599,402.000 & 1,170,744.000 & 412,934.200 \\ 
\hline 
\hline \\[-1.8ex] 
\textit{Note:}  & \multicolumn{4}{r}{$^{*}$p$<$0.1; $^{**}$p$<$0.05; $^{***}$p$<$0.01} \\ 
\end{tabular} }
\end{table}  

Regression of \mrCE{} for the interaction of Age and Skin Type for the CCD dataset can be found in Table~\ref{tbl:comp_int_ccd_5}.

\begin{table}[!htbp] \centering 
  \caption{Interaction (odds ratio) of Age and Skin Type for the CCD dataset.} 
  \label{tbl:comp_int_ccd_5} 
\resizebox{\textwidth}{!}{
\begin{tabular}{@{\extracolsep{5pt}}lcccc} 
\\[-1.8ex]\hline 
\hline \\[-1.8ex] 
 & \multicolumn{4}{c}{\textit{Dependent variable:}} \\ 
\cline{2-5} 
\\[-1.8ex] & \multicolumn{4}{c}{rCE} \\ 
 & Interaction & Just 19-45 & Just 45-64 & Just 65+ \\ 
\\[-1.8ex] & (1) & (2) & (3) & (4)\\ 
\hline \\[-1.8ex] 
 Age45-64 & 1.133 &  &  &  \\ 
  & t = 24.306$^{***}$ &  &  &  \\ 
  & & & & \\ 
 Age65+ & 1.229 &  &  &  \\ 
  & t = 31.779$^{***}$ &  &  &  \\ 
  & & & & \\ 
 FitzDark Fitz & 1.300 & 1.300 & 1.071 & 1.047 \\ 
  & t = 59.191$^{***}$ & t = 59.191$^{***}$ & t = 13.846$^{***}$ & t = 5.506$^{***}$ \\ 
  & & & & \\ 
 Age45-64:FitzDark Fitz & 0.824 &  &  &  \\ 
  & t = $-$28.986$^{***}$ &  &  &  \\ 
  & & & & \\ 
 Age65+:FitzDark Fitz & 0.806 &  &  &  \\ 
  & t = $-$22.705$^{***}$ &  &  &  \\ 
  & & & & \\ 
 Constant & 0.096 & 0.096 & 0.109 & 0.118 \\ 
  & t = $-$659.001$^{***}$ & t = $-$659.001$^{***}$ & t = $-$600.022$^{***}$ & t = $-$394.225$^{***}$ \\ 
  & & & & \\ 
\hline \\[-1.8ex] 
Observations & 4,824,900 & 2,435,175 & 1,783,125 & 606,600 \\ 
Log Likelihood & $-$1,590,420.000 & $-$798,039.400 & $-$585,515.600 & $-$206,865.000 \\ 
Akaike Inf. Crit. & 3,180,852.000 & 1,596,083.000 & 1,171,035.000 & 413,734.000 \\ 
\hline 
\hline \\[-1.8ex] 
\textit{Note:}  & \multicolumn{4}{r}{$^{*}$p$<$0.1; $^{**}$p$<$0.05; $^{***}$p$<$0.01} \\ 
\end{tabular} }
\end{table}  

Regression of \mrCE{} for the interaction of Age, Skin Type, and Gender for the CCD dataset can be found in Table~\ref{tbl:comp_int_ccd_4.asg}.

\begin{table}[!htbp] \centering 
  \caption{Interaction (odds ratio) of Age, Skin Type, and Gender for the CCD dataset.} 
  \label{tbl:comp_int_ccd_4.asg} 
\resizebox{!}{11cm}{
\begin{tabular}{@{\extracolsep{5pt}}lc} 
\\[-1.8ex]\hline 
\hline \\[-1.8ex] 
 & \multicolumn{1}{c}{\textit{Dependent variable:}} \\ 
\cline{2-2} 
\\[-1.8ex] & rCE \\ 
\hline \\[-1.8ex] 
 Age45-64 & 1.072 \\ 
  & t = 9.532$^{***}$ \\ 
  & \\ 
 Age65+ & 1.112 \\ 
  & t = 11.247$^{***}$ \\ 
  & \\ 
 FitzDark Fitz & 1.224 \\ 
  & t = 31.723$^{***}$ \\ 
  & \\ 
 GenderMale & 0.951 \\ 
  & t = $-$6.830$^{***}$ \\ 
  & \\ 
 GenderOther & 0.914 \\ 
  & t = $-$5.774$^{***}$ \\ 
  & \\ 
 Age45-64:FitzDark Fitz & 0.845 \\ 
  & t = $-$17.524$^{***}$ \\ 
  & \\ 
 Age65+:FitzDark Fitz & 0.854 \\ 
  & t = $-$11.784$^{***}$ \\ 
  & \\ 
 Age45-64:GenderMale & 1.116 \\ 
  & t = 10.457$^{***}$ \\ 
  & \\ 
 Age65+:GenderMale & 1.206 \\ 
  & t = 14.308$^{***}$ \\ 
  & \\ 
 Age45-64:GenderOther & 0.940 \\ 
  & t = $-$1.777$^{*}$ \\ 
  & \\ 
 Age65+:GenderOther & 3.348 \\ 
  & t = 28.779$^{***}$ \\ 
  & \\ 
 FitzDark Fitz:GenderMale & 1.141 \\ 
  & t = 14.414$^{***}$ \\ 
  & \\ 
 FitzDark Fitz:GenderOther & 0.980 \\ 
  & t = $-$1.046 \\ 
  & \\ 
 Age45-64:FitzDark Fitz:GenderMale & 0.951 \\ 
  & t = $-$3.665$^{***}$ \\ 
  & \\ 
 Age65+:FitzDark Fitz:GenderMale & 0.868 \\ 
  & t = $-$7.318$^{***}$ \\ 
  & \\ 
 Age45-64:FitzDark Fitz:GenderOther & 1.108 \\ 
  & t = 2.521$^{**}$ \\ 
  & \\ 
 Age65+:FitzDark Fitz:GenderOther &  \\ 
  &  \\ 
  & \\ 
 Constant & 0.099 \\ 
  & t = $-$459.672$^{***}$ \\ 
  & \\ 
\hline \\[-1.8ex] 
Observations & 4,824,900 \\ 
Log Likelihood & $-$1,589,472.000 \\ 
Akaike Inf. Crit. & 3,178,977.000 \\ 
\hline 
\hline \\[-1.8ex] 
\textit{Note:}  & \multicolumn{1}{r}{$^{*}$p$<$0.1; $^{**}$p$<$0.05; $^{***}$p$<$0.01} \\ 
\end{tabular} }
\end{table}  

Regression of \mrCE{} for the interaction of Age (as a numeric variable) and Gender for the CCD dataset can be found in Table~\ref{tbl:comp_int_ccd_4.n}.

\begin{table}[!htbp] \centering 
  \caption{Interaction (odds ratio) of Age (as a numeric variable) and Gender for the CCD dataset.} 
  \label{tbl:comp_int_ccd_4.n} 
\begin{tabular}{@{\extracolsep{5pt}}lc} 
\\[-1.8ex]\hline 
\hline \\[-1.8ex] 
 & \multicolumn{1}{c}{\textit{Dependent variable:}} \\ 
\cline{2-2} 
\\[-1.8ex] & rCE \\ 
\hline \\[-1.8ex] 
 Age\_Numeric & 1.000 \\ 
  & t = 0.787 \\ 
  & \\ 
 GenderMale & 0.954 \\ 
  & t = $-$5.376$^{***}$ \\ 
  & \\ 
 GenderOther & 0.879 \\ 
  & t = $-$12.422$^{***}$ \\ 
  & \\ 
 Age\_Numeric:GenderMale & 1.003 \\ 
  & t = 14.749$^{***}$ \\ 
  & \\ 
 Age\_Numeric:GenderOther & 1.008 \\ 
  & t = 19.447$^{***}$ \\ 
  & \\ 
 Constant & 0.110 \\ 
  & t = $-$356.931$^{***}$ \\ 
  & \\ 
\hline \\[-1.8ex] 
Observations & 4,824,900 \\ 
Log Likelihood & $-$1,591,602.000 \\ 
Akaike Inf. Crit. & 3,183,215.000 \\ 
\hline 
\hline \\[-1.8ex] 
\textit{Note:}  & \multicolumn{1}{r}{$^{*}$p$<$0.1; $^{**}$p$<$0.05; $^{***}$p$<$0.01} \\ 
\end{tabular} 
\end{table}  

Regression of \mrCE{} for the interaction of Age (as a numeric variable) and Skin Type for the CCD dataset can be found in Table~\ref{tbl:comp_int_ccd_5.n}.

\begin{table}[!htbp] \centering 
  \caption{Interaction (odds ratio) of Age (as a numeric variable) and Skin Type for the CCD dataset.} 
  \label{tbl:comp_int_ccd_5.n} 
\begin{tabular}{@{\extracolsep{5pt}}lc} 
\\[-1.8ex]\hline 
\hline \\[-1.8ex] 
 & \multicolumn{1}{c}{\textit{Dependent variable:}} \\ 
\cline{2-2} 
\\[-1.8ex] & rCE \\ 
\hline \\[-1.8ex] 
 Age\_Numeric & 1.005 \\ 
  & t = 35.652$^{***}$ \\ 
  & \\ 
 FitzDark Fitz & 1.385 \\ 
  & t = 40.931$^{***}$ \\ 
  & \\ 
 Age\_Numeric:FitzDark Fitz & 0.996 \\ 
  & t = $-$22.385$^{***}$ \\ 
  & \\ 
 Constant & 0.085 \\ 
  & t = $-$398.455$^{***}$ \\ 
  & \\ 
\hline \\[-1.8ex] 
Observations & 4,824,900 \\ 
Log Likelihood & $-$1,590,484.000 \\ 
Akaike Inf. Crit. & 3,180,977.000 \\ 
\hline 
\hline \\[-1.8ex] 
\textit{Note:}  & \multicolumn{1}{r}{$^{*}$p$<$0.1; $^{**}$p$<$0.05; $^{***}$p$<$0.01} \\ 
\end{tabular} 
\end{table}  

Regression of \mrCE{} for the interaction of Gender and Skin Type for the CCD dataset can be found in Table~\ref{tbl:comp_int_ccd_6}.

\begin{table}[!htbp] \centering 
  \caption{Interaction (odds ratio) of Gender and Skin Type for the CCD dataset.} 
  \label{tbl:comp_int_ccd_6} 
\resizebox{\textwidth}{!}{
\begin{tabular}{@{\extracolsep{5pt}}lcccc} 
\\[-1.8ex]\hline 
\hline \\[-1.8ex] 
 & \multicolumn{4}{c}{\textit{Dependent variable:}} \\ 
\cline{2-5} 
\\[-1.8ex] & \multicolumn{4}{c}{rCE} \\ 
 & Interaction & Just Male & Just Female & Just Other \\ 
\\[-1.8ex] & (1) & (2) & (3) & (4)\\ 
\hline \\[-1.8ex] 
 GenderMale & 1.033 &  &  &  \\ 
  & t = 6.883$^{***}$ &  &  &  \\ 
  & & & & \\ 
 GenderOther & 0.875 &  &  &  \\ 
  & t = $-$9.746$^{***}$ &  &  &  \\ 
  & & & & \\ 
 FitzDark Fitz & 1.123 & 1.213 & 1.123 & 1.228 \\ 
  & t = 26.434$^{***}$ & t = 43.854$^{***}$ & t = 26.431$^{***}$ & t = 12.914$^{***}$ \\ 
  & & & & \\ 
 GenderMale:FitzDark Fitz & 1.080 &  &  &  \\ 
  & t = 12.364$^{***}$ &  &  &  \\ 
  & & & & \\ 
 GenderOther:FitzDark Fitz & 1.094 &  &  &  \\ 
  & t = 5.436$^{***}$ &  &  &  \\ 
  & & & & \\ 
 Constant & 0.103 & 0.107 & 0.103 & 0.090 \\ 
  & t = $-$683.043$^{***}$ & t = $-$672.156$^{***}$ & t = $-$682.936$^{***}$ & t = $-$181.136$^{***}$ \\ 
  & & & & \\ 
\hline \\[-1.8ex] 
Observations & 4,824,900 & 2,232,225 & 2,375,325 & 217,350 \\ 
Log Likelihood & $-$1,590,636.000 & $-$754,509.000 & $-$768,396.800 & $-$67,730.120 \\ 
Akaike Inf. Crit. & 3,181,284.000 & 1,509,022.000 & 1,536,798.000 & 135,464.200 \\ 
\hline 
\hline \\[-1.8ex] 
\textit{Note:}  & \multicolumn{4}{r}{$^{*}$p$<$0.1; $^{**}$p$<$0.05; $^{***}$p$<$0.01} \\ 
\end{tabular} }
\end{table}

\subsubsection{MIAP Demographic Interactions}\label{sec:comp_int_miap}
Regression of \mrCE{} for the interaction of Age and Gender for the MIAP dataset can be found in Table~\ref{tbl:comp_int_miap_1}.

\begin{table}[!htbp] \centering 
  \caption{Interaction (odds ratio) of Age and Gender for the MIAP dataset.} 
  \label{tbl:comp_int_miap_1} 
\resizebox{\textwidth}{!}{\begin{tabular}{@{\extracolsep{5pt}}lccccccc} 
\\[-1.8ex]\hline 
\hline \\[-1.8ex] 
 & \multicolumn{7}{c}{\textit{Dependent variable:}} \\ 
\cline{2-8} 
\\[-1.8ex] & \multicolumn{7}{c}{rCE} \\ 
 & Interaction & Just Young & Just Middle & Just Old & Just Feminine & Just Masculine & Just Unknown \\ 
\\[-1.8ex] & (1) & (2) & (3) & (4) & (5) & (6) & (7)\\
\end{tabular} }
\end{table}  

\subsubsection{UTKFace Demographic Interactions}\label{sec:comp_int_utk}
Regression of \mrCE{} for the interaction of Age and Gender for the UTKFace dataset can be found in Table~\ref{tbl:comp_int_utk_1}.

\begin{table}[!htbp] \centering 
  \caption{Interaction (odds ratio) of Age and Gender for the UTKFace dataset.} 
  \label{tbl:comp_int_utk_1} 
\resizebox{\textwidth}{!}{
\begin{tabular}{@{\extracolsep{5pt}}lccccc} 
\\[-1.8ex]\hline 
\hline \\[-1.8ex] 
 & \multicolumn{5}{c}{\textit{Dependent variable:}} \\ 
\cline{2-6} 
\\[-1.8ex] & \multicolumn{5}{c}{$rCE$} \\ 
 & Interaction & Just 0-18 & Just 19-45 & Just 45-64 & Just 65+ \\ 
\\[-1.8ex] & (1) & (2) & (3) & (4) & (5)\\ 
\hline \\[-1.8ex] 
 Age19-45 & 0.855 &  &  &  &  \\ 
  & t = $-$22.893$^{***}$ &  &  &  &  \\ 
  & & & & & \\ 
 Age45-64 & 0.964 &  &  &  &  \\ 
  & t = $-$3.723$^{***}$ &  &  &  &  \\ 
  & & & & & \\ 
 Age65+ & 1.287 &  &  &  &  \\ 
  & t = 25.011$^{***}$ &  &  &  &  \\ 
  & & & & & \\ 
 genderMale & 0.945 & 0.945 & 1.009 & 1.036 & 0.914 \\ 
  & t = $-$6.807$^{***}$ & t = $-$6.807$^{***}$ & t = 1.710$^{*}$ & t = 3.620$^{***}$ & t = $-$7.823$^{***}$ \\ 
  & & & & & \\ 
 Age19-45:genderMale & 1.069 &  &  &  &  \\ 
  & t = 6.625$^{***}$ &  &  &  &  \\ 
  & & & & & \\ 
 Age45-64:genderMale & 1.096 &  &  &  &  \\ 
  & t = 7.198$^{***}$ &  &  &  &  \\ 
  & & & & & \\ 
 Age65+:genderMale & 0.968 &  &  &  &  \\ 
  & t = $-$2.298$^{**}$ &  &  &  &  \\ 
  & & & & & \\ 
 Constant & 0.105 & 0.105 & 0.090 & 0.101 & 0.135 \\ 
  & t = $-$395.638$^{***}$ & t = $-$395.638$^{***}$ & t = $-$631.394$^{***}$ & t = $-$282.362$^{***}$ & t = $-$239.992$^{***}$ \\ 
  & & & & & \\ 
\hline \\[-1.8ex] 
Observations & 3,314,387 & 678,662 & 1,728,150 & 606,600 & 300,975 \\ 
Log Likelihood & $-$1,000,139.000 & $-$209,933.600 & $-$493,906.800 & $-$189,325.500 & $-$106,973.500 \\ 
Akaike Inf. Crit. & 2,000,295.000 & 419,871.200 & 987,817.600 & 378,654.900 & 213,951.000 \\ 
\hline 
\hline \\[-1.8ex] 
\textit{Note:}  & \multicolumn{5}{r}{$^{*}$p$<$0.1; $^{**}$p$<$0.05; $^{***}$p$<$0.01} \\ 
\end{tabular} }
\end{table}  

Regression of \mrCE{} for the interaction of Age and Ethnicity for the UTKFace dataset can be found in Table~\ref{tbl:comp_int_utk_2}.

\begin{table}[!htbp] \centering 
  \caption{Interaction (odds ratio) of Age and Ethnicity for the UTKFace dataset.} 
  \label{tbl:comp_int_utk_2} 
\resizebox{\textwidth}{11cm}{
\begin{tabular}{@{\extracolsep{5pt}}lccccc} 
\\[-1.8ex]\hline 
\hline \\[-1.8ex] 
 & \multicolumn{5}{c}{\textit{Dependent variable:}} \\ 
\cline{2-6} 
\\[-1.8ex] & \multicolumn{5}{c}{$rCE$} \\ 
 & Interaction & Just 0-18 & Just 19-45 & Just 45-64 & Just 65+ \\ 
\\[-1.8ex] & (1) & (2) & (3) & (4) & (5)\\
\end{tabular} }
\end{table}  

Regression of \mrCE{} for the interaction of Gender and Ethnicity for the UTKFace dataset can be found in Table~\ref{tbl:comp_int_utk_3}.

Regression of \mrCE{} for the interaction of Gender and Ethnicity for the UTKFace dataset can be found in Table~\ref{tbl:comp_int_utk_3}.

\begin{table}[!htbp] \centering 
  \caption{Interaction (odds ratio) of Gender and Ethnicity for the UTKFace dataset.} 
  \label{tbl:comp_int_utk_3} 
\begin{tabular}{@{\extracolsep{5pt}}lccc} 
\\[-1.8ex]\hline 
\hline \\[-1.8ex] 
 & \multicolumn{3}{c}{\textit{Dependent variable:}} \\ 
\cline{2-4} 
\\[-1.8ex] & \multicolumn{3}{c}{$rCE$} \\ 
 & Interaction & Just Male & Just Female \\ 
\\[-1.8ex] & (1) & (2) & (3)\\ 
\hline \\[-1.8ex] 
 genderMale & 0.822 &  &  \\ 
  & t = $-$13.125$^{***}$ &  &  \\ 
  & & & \\ 
 ethnicityWhite & 1.198 & 1.420 & 1.198 \\ 
  & t = 17.206$^{***}$ & t = 28.782$^{***}$ & t = 17.206$^{***}$ \\ 
  & & & \\ 
 ethnicityBlack & 1.074 & 1.424 & 1.074 \\ 
  & t = 6.190$^{***}$ & t = 27.375$^{***}$ & t = 6.190$^{***}$ \\ 
  & & & \\ 
 ethnicityAsian & 0.957 & 1.251 & 0.957 \\ 
  & t = $-$3.732$^{***}$ & t = 16.519$^{***}$ & t = $-$3.732$^{***}$ \\ 
  & & & \\ 
 ethnicityIndian & 0.908 & 1.084 & 0.908 \\ 
  & t = $-$8.045$^{***}$ & t = 6.121$^{***}$ & t = $-$8.045$^{***}$ \\ 
  & & & \\ 
 genderMale:ethnicityWhite & 1.186 &  &  \\ 
  & t = 10.601$^{***}$ &  &  \\ 
  & & & \\ 
 genderMale:ethnicityBlack & 1.326 &  &  \\ 
  & t = 16.291$^{***}$ &  &  \\ 
  & & & \\ 
 genderMale:ethnicityAsian & 1.306 &  &  \\ 
  & t = 14.934$^{***}$ &  &  \\ 
  & & & \\ 
 genderMale:ethnicityIndian & 1.194 &  &  \\ 
  & t = 9.945$^{***}$ &  &  \\ 
  & & & \\ 
 Constant & 0.093 & 0.076 & 0.093 \\ 
  & t = $-$249.112$^{***}$ & t = $-$224.112$^{***}$ & t = $-$249.112$^{***}$ \\ 
  & & & \\ 
\hline \\[-1.8ex] 
Observations & 3,314,387 & 1,729,045 & 1,585,342 \\ 
Log Likelihood & $-$1,000,032.000 & $-$522,244.200 & $-$477,787.900 \\ 
Akaike Inf. Crit. & 2,000,084.000 & 1,044,498.000 & 955,585.700 \\ 
\hline 
\hline \\[-1.8ex] 
\textit{Note:}  & \multicolumn{3}{r}{$^{*}$p$<$0.1; $^{**}$p$<$0.05; $^{***}$p$<$0.01} \\ 
\end{tabular} 
\end{table}  
\section{Statistical Significance Regressions for Gender Prediction}\label{sec:gender_pred}

Regression of gender estimation for all variables for the all the datasets with unified demographic variables can be found in Table~\ref{tbl:comp_gender_full}.

\begin{table}[!htbp] \centering 
  \caption{Gender estimation (odds ratio) for the all datasets with unifed demographic variables.} 
  \label{tbl:comp_gender_full} 
\resizebox{!}{11cm}{
\begin{tabular}{@{\extracolsep{5pt}}lcccc} 
\\[-1.8ex]\hline 
\hline \\[-1.8ex] 
 & \multicolumn{4}{c}{\textit{Dependent variable:}} \\ 
\cline{2-5} 
\\[-1.8ex] & \multicolumn{4}{c}{pred\_gender\_iswrong} \\ 
 & Adience & CCD & MIAP & UTKFace \\ 
\\[-1.8ex] & (1) & (2) & (3) & (4)\\
\end{tabular} }
\end{table}

Regression of gender prediction for all variables for the Adience dataset can be found in Table~\ref{tbl:comp_gender_adience}.

\begin{table}[!htbp] \centering 
  \caption{Gender prediction (odds ratio) for the Adience dataset.} 
  \label{tbl:comp_gender_adience} 
\resizebox{!}{11cm}{
\begin{tabular}{@{\extracolsep{5pt}}lc} 
\\[-1.8ex]\hline 
\hline \\[-1.8ex] 
 & \multicolumn{1}{c}{\textit{Dependent variable:}} \\ 
\cline{2-2} 
\\[-1.8ex] & pred\_gender\_iswrong \\ 
\hline \\[-1.8ex] 
 Age3-7 & 1.092 \\ 
  & t = 8.348$^{***}$ \\
\end{tabular} }
\end{table}  

Regression of gender estimation for all variables for the CCD dataset can be found in Table~\ref{tbl:comp_gender_ccd}.

\begin{table}[!htbp] \centering 
  \caption{Gender estimation (odds ratio) for the CCD dataset.} 
  \label{tbl:comp_gender_ccd} 
\resizebox{!}{11cm}{
\begin{tabular}{@{\extracolsep{5pt}}lc} 
\\[-1.8ex]\hline 
\hline \\[-1.8ex] 
 & \multicolumn{1}{c}{\textit{Dependent variable:}} \\ 
\cline{2-2} 
\\[-1.8ex] & pred\_gender\_iswrong \\ 
\hline \\[-1.8ex] 
 Age\_Numeric & 1.000 \\ 
  & t = $-$3.126$^{***}$ \\ 
  & \\ 
 GenderMale & 1.114 \\ 
  & t = 23.716$^{***}$ \\ 
  & \\ 
 FitzDark Fitz & 1.073 \\ 
  & t = 14.978$^{***}$ \\ 
  & \\ 
 lightingDark & 1.342 \\ 
  & t = 58.560$^{***}$ \\ 
  & \\ 
 corruptiongaussian-noise & 7.919 \\ 
  & t = 62.206$^{***}$ \\ 
  & \\ 
 corruptionshot-noise & 6.794 \\ 
  & t = 57.512$^{***}$ \\ 
  & \\ 
 corruptionimpulse-noise & 7.898 \\ 
  & t = 62.123$^{***}$ \\ 
  & \\ 
 corruptiondefocus-blur & 1.293 \\ 
  & t = 7.345$^{***}$ \\ 
  & \\ 
 corruptionglass-blur & 1.027 \\ 
  & t = 0.747 \\ 
  & \\ 
 corruptionmotion-blur & 1.236 \\ 
  & t = 6.038$^{***}$ \\ 
  & \\ 
 corruptionzoom-blur & 29.170 \\ 
  & t = 101.732$^{***}$ \\ 
  & \\ 
 corruptionsnow & 7.490 \\ 
  & t = 60.500$^{***}$ \\ 
  & \\ 
 corruptionfrost & 7.383 \\ 
  & t = 60.058$^{***}$ \\ 
  & \\ 
 corruptionfog & 1.856 \\ 
  & t = 17.983$^{***}$ \\ 
  & \\ 
 corruptionbrightness & 1.177 \\ 
  & t = 4.627$^{***}$ \\ 
  & \\ 
 corruptioncontrast & 4.522 \\ 
  & t = 45.031$^{***}$ \\ 
  & \\ 
 corruptionelastic-transform & 1.122 \\ 
  & t = 3.269$^{***}$ \\ 
  & \\ 
 corruptionpixelate & 1.815 \\ 
  & t = 17.311$^{***}$ \\ 
  & \\ 
 corruptionjpeg-compression & 1.440 \\ 
  & t = 10.471$^{***}$ \\ 
  & \\ 
 Constant & 0.043 \\ 
  & t = $-$94.100$^{***}$ \\ 
  & \\ 
\hline \\[-1.8ex] 
Observations & 1,556,328 \\ 
Log Likelihood & $-$617,424.800 \\ 
Akaike Inf. Crit. & 1,234,890.000 \\ 
\hline 
\hline \\[-1.8ex] 
\textit{Note:}  & \multicolumn{1}{r}{$^{*}$p$<$0.1; $^{**}$p$<$0.05; $^{***}$p$<$0.01} \\ 
\end{tabular} }
\end{table}

Regression of gender estimation for all variables for the MIAP dataset can be found in Table~\ref{tbl:comp_gender_miap}.

\begin{table}[!htbp] \centering 
  \caption{Gender estimation (odds ratio) for the MIAP dataset.} 
  \label{tbl:comp_gender_miap} 
\resizebox{!}{11cm}{
\begin{tabular}{@{\extracolsep{5pt}}lc} 
\\[-1.8ex]\hline 
\hline \\[-1.8ex] 
 & \multicolumn{1}{c}{\textit{Dependent variable:}} \\ 
\cline{2-2} 
\\[-1.8ex] & pred\_gender\_iswrong \\ 
\hline \\[-1.8ex] 
 AgePresentationOlder & 0.696 \\ 
  & t = $-$36.178$^{***}$ \\ 
  & \\ 
 AgePresentationUnknown & 8.443 \\ 
  & t = 174.537$^{***}$ \\ 
  & \\ 
 GenderPresentationPredominantly Masculine & 1.121 \\ 
  & t = 14.530$^{***}$ \\ 
  & \\ 
 corruptiongaussian-noise & 2.790 \\ 
  & t = 25.488$^{***}$ \\ 
  & \\ 
 corruptionshot-noise & 2.847 \\ 
  & t = 25.995$^{***}$ \\ 
  & \\ 
 corruptionimpulse-noise & 2.914 \\ 
  & t = 26.580$^{***}$ \\ 
  & \\ 
 corruptiondefocus-blur & 1.840 \\ 
  & t = 15.043$^{***}$ \\ 
  & \\ 
 corruptionglass-blur & 1.218 \\ 
  & t = 4.822$^{***}$ \\ 
  & \\ 
 corruptionmotion-blur & 1.965 \\ 
  & t = 16.689$^{***}$ \\ 
  & \\ 
 corruptionzoom-blur & 6.835 \\ 
  & t = 47.732$^{***}$ \\ 
  & \\ 
 corruptionsnow & 2.537 \\ 
  & t = 23.100$^{***}$ \\ 
  & \\ 
 corruptionfrost & 2.639 \\ 
  & t = 24.091$^{***}$ \\ 
  & \\ 
 corruptionfog & 1.576 \\ 
  & t = 11.191$^{***}$ \\ 
  & \\ 
 corruptionbrightness & 1.264 \\ 
  & t = 5.717$^{***}$ \\ 
  & \\ 
 corruptioncontrast & 2.835 \\ 
  & t = 25.890$^{***}$ \\ 
  & \\ 
 corruptionelastic-transform & 1.347 \\ 
  & t = 7.295$^{***}$ \\ 
  & \\ 
 corruptionpixelate & 2.045 \\ 
  & t = 17.690$^{***}$ \\ 
  & \\ 
 corruptionjpeg-compression & 1.421 \\ 
  & t = 8.621$^{***}$ \\ 
  & \\ 
 Constant & 0.209 \\ 
  & t = $-$41.232$^{***}$ \\ 
  & \\ 
\hline \\[-1.8ex] 
Observations & 350,284 \\ 
Log Likelihood & $-$204,156.500 \\ 
Akaike Inf. Crit. & 408,350.900 \\ 
\hline 
\hline \\[-1.8ex] 
\textit{Note:}  & \multicolumn{1}{r}{$^{*}$p$<$0.1; $^{**}$p$<$0.05; $^{***}$p$<$0.01} \\ 
\end{tabular} }
\end{table}  

Regression of gender estimation for all variables for the UTKFace dataset can be found in Table~\ref{tbl:comp_gender_utk}.

\begin{table}[!htbp] \centering 
  \caption{Gender estimation (odds ratio) for the UTKFace dataset.} 
  \label{tbl:comp_gender_utk} 
\resizebox{!}{11cm}{
\begin{tabular}{@{\extracolsep{5pt}}lc} 
\\[-1.8ex]\hline 
\hline \\[-1.8ex] 
 & \multicolumn{1}{c}{\textit{Dependent variable:}} \\ 
\cline{2-2} 
\\[-1.8ex] & pred\_gender\_iswrong \\ 
\hline \\[-1.8ex] 
 Age\_Numeric & 0.974 \\ 
  & t = $-$225.620$^{***}$ \\ 
  & \\ 
 GenderMale & 2.804 \\ 
  & t = 225.578$^{***}$ \\ 
  & \\ 
 corruptiongaussian-noise & 6.064 \\ 
  & t = 66.509$^{***}$ \\ 
  & \\ 
 corruptionshot-noise & 6.976 \\ 
  & t = 71.769$^{***}$ \\ 
  & \\ 
 corruptionimpulse-noise & 5.699 \\ 
  & t = 64.174$^{***}$ \\ 
  & \\ 
 corruptiondefocus-blur & 2.931 \\ 
  & t = 39.199$^{***}$ \\ 
  & \\ 
 corruptionglass-blur & 2.940 \\ 
  & t = 39.321$^{***}$ \\ 
  & \\ 
 corruptionmotion-blur & 2.640 \\ 
  & t = 35.305$^{***}$ \\ 
  & \\ 
 corruptionzoom-blur & 1.971 \\ 
  & t = 24.444$^{***}$ \\ 
  & \\ 
 corruptionsnow & 1.921 \\ 
  & t = 23.503$^{***}$ \\ 
  & \\ 
 corruptionfrost & 2.839 \\ 
  & t = 38.018$^{***}$ \\ 
  & \\ 
 corruptionfog & 1.728 \\ 
  & t = 19.618$^{***}$ \\ 
  & \\ 
 corruptionbrightness & 1.172 \\ 
  & t = 5.574$^{***}$ \\ 
  & \\ 
 corruptioncontrast & 2.722 \\ 
  & t = 36.434$^{***}$ \\ 
  & \\ 
 corruptionelastic-transform & 2.437 \\ 
  & t = 32.321$^{***}$ \\ 
  & \\ 
 corruptionpixelate & 1.436 \\ 
  & t = 12.862$^{***}$ \\ 
  & \\ 
 corruptionjpeg-compression & 1.603 \\ 
  & t = 16.853$^{***}$ \\ 
  & \\ 
 Constant & 0.091 \\ 
  & t = $-$90.381$^{***}$ \\ 
  & \\ 
\hline \\[-1.8ex] 
Observations & 1,679,524 \\ 
Log Likelihood & $-$698,055.600 \\ 
Akaike Inf. Crit. & 1,396,147.000 \\ 
\hline 
\hline \\[-1.8ex] 
\textit{Note:}  & \multicolumn{1}{r}{$^{*}$p$<$0.1; $^{**}$p$<$0.05; $^{***}$p$<$0.01} \\ 
\end{tabular} }
\end{table}

\section{Statistical Significance Regressions for Age Estimation}\label{sec:age_est}

Regression of Age estimation for all variables for the CCD dataset can be found in Table~\ref{tbl:comp_age_ccd}.

\begin{table}[!htbp] \centering 
  \caption{Age estimation for the CCD dataset.} 
  \label{tbl:comp_age_ccd} 
\resizebox{!}{11cm}{\begin{tabular}{@{\extracolsep{5pt}}lc} 
\\[-1.8ex]\hline 
\hline \\[-1.8ex] 
 & \multicolumn{1}{c}{\textit{Dependent variable:}} \\ 
\cline{2-2} 
\\[-1.8ex] & diff\_pred\_age \\ 
\hline \\[-1.8ex] 
 Age\_Numeric & 0.634$^{***}$ \\ 
  & (0.001) \\ 
  & \\ 
 GenderMale & $-$2.866$^{***}$ \\ 
  & (0.017) \\ 
  & \\ 
 GenderOther & $-$11.029$^{***}$ \\ 
  & (0.044) \\ 
  & \\ 
 FitzDark Fitz & 1.869$^{***}$ \\ 
  & (0.016) \\ 
  & \\ 
 lightingDark & 0.697$^{***}$ \\ 
  & (0.019) \\ 
  & \\ 
 corruptiongaussian-noise & 6.525$^{***}$ \\ 
  & (0.075) \\ 
  & \\ 
 corruptionshot-noise & 4.967$^{***}$ \\ 
  & (0.075) \\ 
  & \\ 
 corruptionimpulse-noise & 6.646$^{***}$ \\ 
  & (0.075) \\ 
  & \\ 
 corruptiondefocus-blur & 0.528$^{***}$ \\ 
  & (0.075) \\ 
  & \\ 
 corruptionglass-blur & 0.361$^{***}$ \\ 
  & (0.075) \\ 
  & \\ 
 corruptionmotion-blur & 1.569$^{***}$ \\ 
  & (0.075) \\ 
  & \\ 
 corruptionzoom-blur & 4.706$^{***}$ \\ 
  & (0.080) \\ 
  & \\ 
 corruptionsnow & 2.723$^{***}$ \\ 
  & (0.075) \\ 
  & \\ 
 corruptionfrost & 0.824$^{***}$ \\ 
  & (0.076) \\ 
  & \\ 
 corruptionfog & 3.288$^{***}$ \\ 
  & (0.075) \\ 
  & \\ 
 corruptionbrightness & 0.863$^{***}$ \\ 
  & (0.075) \\ 
  & \\ 
 corruptioncontrast & 4.763$^{***}$ \\ 
  & (0.075) \\ 
  & \\ 
 corruptionelastic-transform & $-$0.482$^{***}$ \\ 
  & (0.075) \\ 
  & \\ 
 corruptionpixelate & 0.622$^{***}$ \\ 
  & (0.075) \\ 
  & \\ 
 corruptionjpeg-compression & $-$0.942$^{***}$ \\ 
  & (0.075) \\ 
  & \\ 
 Constant & $-$20.136$^{***}$ \\ 
  & (0.073) \\ 
  & \\ 
\hline \\[-1.8ex] 
Observations & 1,538,072 \\ 
Log Likelihood & $-$5,723,810.000 \\ 
Akaike Inf. Crit. & 11,447,661.000 \\ 
\hline 
\hline \\[-1.8ex] 
\textit{Note:}  & \multicolumn{1}{r}{$^{*}$p$<$0.1; $^{**}$p$<$0.05; $^{***}$p$<$0.01} \\ 
\end{tabular} }
\end{table}  

Regression of age estimation for all variables for the UTKFace dataset can be found in Table~\ref{tbl:comp_age_utk}.

\begin{table}[!htbp] \centering 
  \caption{Age estimation for the UTKFace dataset.} 
  \label{tbl:comp_age_utk} 
\resizebox{!}{11cm}{\begin{tabular}{@{\extracolsep{5pt}}lc} 
\\[-1.8ex]\hline 
\hline \\[-1.8ex] 
 & \multicolumn{1}{c}{\textit{Dependent variable:}} \\ 
\cline{2-2} 
\\[-1.8ex] & pred\_gender\_iswrong \\ 
\hline \\[-1.8ex] 
 Age\_Numeric & $-$0.004$^{***}$ \\ 
  & (0.00001) \\ 
  & \\ 
 GenderMale & 0.135$^{***}$ \\ 
  & (0.001) \\ 
  & \\ 
 corruptiongaussian-noise & 0.091$^{***}$ \\ 
  & (0.002) \\ 
  & \\ 
 corruptionshot-noise & 0.107$^{***}$ \\ 
  & (0.002) \\ 
  & \\ 
 corruptionimpulse-noise & 0.113$^{***}$ \\ 
  & (0.002) \\ 
  & \\ 
 corruptiondefocus-blur & 0.090$^{***}$ \\ 
  & (0.002) \\ 
  & \\ 
 corruptionglass-blur & 0.071$^{***}$ \\ 
  & (0.002) \\ 
  & \\ 
 corruptionmotion-blur & 0.052$^{***}$ \\ 
  & (0.002) \\ 
  & \\ 
 corruptionzoom-blur & 0.041$^{***}$ \\ 
  & (0.002) \\ 
  & \\ 
 corruptionsnow & 0.051$^{***}$ \\ 
  & (0.002) \\ 
  & \\ 
 corruptionfrost & 0.053$^{***}$ \\ 
  & (0.002) \\ 
  & \\ 
 corruptionfog & 0.042$^{***}$ \\ 
  & (0.002) \\ 
  & \\ 
 corruptionbrightness & 0.011$^{***}$ \\ 
  & (0.002) \\ 
  & \\ 
 corruptioncontrast & 0.067$^{***}$ \\ 
  & (0.002) \\ 
  & \\ 
 corruptionelastic-transform & 0.031$^{***}$ \\ 
  & (0.002) \\ 
  & \\ 
 corruptionpixelate & 0.009$^{***}$ \\ 
  & (0.002) \\ 
  & \\ 
 corruptionjpeg-compression & 0.017$^{***}$ \\ 
  & (0.002) \\ 
  & \\ 
 Constant & 0.121$^{***}$ \\ 
  & (0.002) \\ 
  & \\ 
\hline \\[-1.8ex] 
Observations & 1,590,377 \\ 
Log Likelihood & $-$438,654.600 \\ 
Akaike Inf. Crit. & 877,345.200 \\ 
\hline 
\hline \\[-1.8ex] 
\textit{Note:}  & \multicolumn{1}{r}{$^{*}$p$<$0.1; $^{**}$p$<$0.05; $^{***}$p$<$0.01} \\ 
\end{tabular} }
\end{table}

\end{document}